\documentclass[12pt,letterpaper]{article}
\usepackage[utf8]{inputenc}

\usepackage{graphicx,array}
\usepackage{url}
\usepackage{color}
\usepackage{latexsym}
\usepackage{amsthm}
\usepackage{amsmath}
\usepackage{amssymb}
\usepackage{amsfonts}
\usepackage[numbers,sort&compress]{natbib}
\usepackage{bm}
\usepackage{slashed}
\usepackage{mathrsfs}
\usepackage{enumerate}
\usepackage{tikz}
\usepackage{siunitx}
\usepackage{mdframed}
\usepackage{setspace}  
\usepackage{esvect}
\usepackage{caption}

\usepackage{tcolorbox}%

\usepackage{hyperref} 
\hypersetup{
    colorlinks=true,       
    linkcolor=red,          
    citecolor=blue,        
    filecolor=magenta,      
    urlcolor=blue           
}
\usepackage[all]{hypcap} 
\usepackage{multirow}
\usepackage{multicol}
\usepackage{makecell}

\usepackage{natbib}
\newcommand{\nc}{\newcommand}  
\newcommand{\mc}{\mathcal}

\newcommand{\uu}{\;\!}

\nc{\beq}{\begin{equation}}
\nc{\eeq}{\end{equation}}
\nc{\beqa}{\begin{eqnarray}}  
\nc{\eeqa}{\end{eqnarray}}  
\nc{\bit}{\begin{itemize}}  
\nc{\eit}{\end{itemize}}  

\def\GeV{\mathrm{GeV}}     

\newcommand{\eg}{{\it e.g.}}
\newcommand{\ie}{{\it i.e.}}

\newcommand{\mmed}{m_{\rm med}}
\newcommand{\GpoG}{\beta}
\newcommand{\fGW}{f_{\rm GW}}
\newcommand{\fGWs}{f_{{\rm GW},s}}
\newcommand{\fmaxs}{f_{\rm max,s}}
\newcommand{\MDM}{\text{{\bf \tiny MDM}}}
\newcommand{\mdm}{M_\MDM}
\newcommand{\rdm}{R_\MDM}
\newcommand{\rhodm}{\rho_\MDM}

\newcommand{\Rdec}{R_{\rm dec}}

\newcommand{\aemit}{a_{\rm emit}}
\newcommand{\rmax}{r_{\rm max}}
\newcommand{\rmin}{r_{\rm min}}
\newcommand{\fdm}{f_{\rm DM}}
\newcommand{\taumix}{\tau_{\rm mix}}

\newcommand{\Ueff}{U_{\rm eff}}
\newcommand{\Ubarrier}{U_{\rm barrier}}
\newcommand{\Ucross}{U_{\rm cross}}
\newcommand{\GpoGbarrier}{\GpoG_{\rm barrier}}

\title{ 
 {\bf Gravitational Waves From Dark Binaries With Finite-Range Dark Forces
 }\\
\author{\large Yang Bai$^{\,\circ}$, Sida Lu$^{\,\dagger\sharp\footnote{Corresponding author}}$\,, and Nicholas Orlofsky$^{\, \diamond}$}
\date{\small \it 
$^\dagger$School of Physics and Astronomy, Sun Yat-sen University (Zhuhai Campus), Zhuhai 519082, China\\
$^\sharp$Institute for Advanced Study, The Hong Kong University of Science and Technology, \\Clear Water Bay, Kowloon, Hong Kong S.A.R., P. R. China \\
$^\circ$Department of Physics, University of Wisconsin-Madison, Madison, WI 53706, USA\\
$^\diamond$Institute of Theoretical Physics, Faculty of Physics, University of Warsaw, ul. Pasteura 5, PL-02-093 Warsaw, Poland \\
}
}

\begin{document}

\maketitle

\setlength{\parskip}{0.2ex}

\begin{abstract} 
This paper calculates the stochastic gravitational wave background from dark binaries with finite-range attractive dark forces, complementing  previous works which consider long-range dark forces. The finiteness of the dark force range can dramatically modify both the initial distributions and evolution histories of the binaries. The generated gravitational wave spectrum is enhanced in the intermediate frequency regime and exhibits interesting ``knee" and ``ankle" features, the most common of which is related to the turn on of the dark force mediator radiation. Other such spectral features are related to changes in the binary merger lifetime and the probability distribution for the initial binary separation. The stochastic gravitational wave background from sub-solar-mass dark binaries is detectable by both space- and ground-based gravitational wave observatories.
\end{abstract}

\thispagestyle{empty}  
\newpage   
\setcounter{page}{1}  

\begingroup
\hypersetup{linkcolor=black,linktocpage}
\tableofcontents
\endgroup

\newpage

\section{Introduction}
\label{sec:intro}
The success of recent gravitational wave (GW) observatories based on interferometry \cite{LIGOScientific:2014pky,LIGOScientific:2016aoc} and pulsar timing arrays (PTAs) \cite{NANOGrav:2023gor,NANOGrav:2023hde,NANOGrav:2023hvm,NANOGrav:2023hfp,EPTA:2023sfo,EPTA:2023fyk,EPTA:2023gyr,EPTA:2023xxk,Reardon:2023gzh,Reardon:2023zen,Zic:2023gta,Xu:2023wog} provides a new probe for physics beyond the Standard Model (BSM). 
Specifically, due to the extremely weak interaction strength between gravity and matter, the stochastic GW background (SGWB) can serve as a recorder of the cosmic history, offering unique insights into all eras between inflation and today. 
Mergers of compact objects in the Standard Model (SM) are predicted to contribute to the SGWB, including those involving astrophysically formed supermassive black holes that may offer a partial or full explanation for the likely GW signal at PTA experiments \cite{NANOGrav:2023hfp,EPTA:2023xxk,Ellis:2023oxs}.
Possible BSM sources of the SGWB involve 
first-order cosmic phase transitions, cosmic strings, 
domain walls, primordial density perturbations, inflationary tensor perturbations, and in particular the merging binaries of macroscopic dark matter (MDM) objects, such as primordial black holes \cite{Mandic:2016lcn,Wang:2016ana,Raidal:2017mfl,Sasaki:2018dmp,Pujolas:2021yaw,Garcia-Bellido:2021jlq,Atal:2022zux,Braglia:2022icu,Banerjee:2023brn}, fermion solitons~\cite{Lee:1986tr,DelGrosso:2023trq,Xie:2024mxr}, 
scalar solitons~\cite{Friedberg:1976me,Coleman:1985ki,Lee:1991ax,Ponton:2019hux,Nugaev:2019vru,Heeck:2020bau,Bai:2021mzu,Bai:2022kxq}, 
quark nuggets~\cite{Krnjaic:2014xza,Bai:2018vik,Bai:2018dxf,Liang:2016tqc}, dark nuclei~\cite{Wise:2014jva,Wise:2014ola,Hardy:2014mqa,Gresham:2017zqi,Gresham:2017cvl}, or mirror stars~\cite{Curtin:2019lhm,Curtin:2019ngc,Kouvaris:2015rea,Giudice:2016zpa,Maselli:2017vfi,Hippert:2021fch,Gross:2021qgx,Ryan:2022hku,Banks:2023eym,Bai:2023mfi}. 
These MDM binaries form predominantly in the early universe before matter-radiation equality \cite{Nakamura:1997sm,Ioka:1998nz,Sasaki:2016jop,Ali-Haimoud:2017rtz,Raidal:2017mfl,Raidal:2018bbj,Vaskonen:2019jpv,Hutsi:2020sol,Bai:2023lyf}, when the mutual attractive force between nearest-neighbor MDM pairs overcomes the dragging from the Hubble flow. The initial shapes of the binary orbits are determined by the separation between the binaries as well as the tidal forces exerted on them by other nearby MDMs. Then, these binaries inspiral and merge, contributing to the SGWB. It is important to understand such dark sector phenomena that can lead to a SGWB because, if the dark sector has very weak or nonexistent nongravitational couplings with the SM, gravitational probes like this may be the only way to measure the properties of DM in what is otherwise considered a ``nightmare scenario'' for DM detection. 

In general, it is quite possible that a secluded dark sector may contain an array of fields and forces, similar to the SM. This is especially true for dark sectors that admit the formation of MDM objects.
The SGWB from MDM binaries assuming that the dark sector has an additional attractive secluded long-range dark force (DF) was previously studied in \cite{Bai:2023lyf}  (see also Refs.~\cite{Krause:1994ar,Croon:2017zcu,Alexander:2018qzg,Dror:2019uea} for related work with visible sector binaries and new forces).
This additional attractive force binds the binaries more tightly, altering not only their formation but also their evolution after decoupling from the Hubble flow.
As a result, the SGWB generated by the merger of the binaries is influenced by the DF in three ways compared with the gravity-only case: a) the number of decoupled MDM pairs is increased; b) the GW emission spectrum at the source is modified because the GW emission is increased and the DF mediator can also be radiated; and c) the binary lifetime is shortened due to both (b) and the fact that binaries will decouple from the Hubble flow earlier and form tighter initial binaries. Points (a) and (c) tend to enhance the SGWB by increasing the binary merger rate. However, if the binary lifetime becomes too short, the binaries may merge too quickly after formation such that the emitted GWs are redshifted away, suppressing the SGWB. Regarding (b), the GW source emission can be enhanced by the additional attractive interaction, but it can also be suppressed if the radiation of the DF mediator steals away too much energy from the binary. Generally speaking, SGWB enhancement requires an attractive DF that is much stronger than gravity; otherwise, there is very little enhancement to the binary orbital frequency (and thus GW emission) while much of the orbital energy is instead radiated away by the DF mediator. But, the DF cannot be too strong either; otherwise, binaries decouple from the Hubble flow very early, forming with very small initial separation and merging too soon.

Ref.~\cite{Bai:2023lyf} considered only the case where the DF has an approximately massless mediator. Such an assumption is entirely consistent with existing constraints, which only require the DF range to be less than tens of kiloparsecs \cite{Savastano:2019zpr,Bogorad:2023wzn}, well above the sizes of the binaries in question. Nevertheless, as this work explores, there are more opportunities to enhance the SGWB with a (not too) massive DF mediator. A key observation is that the finite mediator mass opens a frequency window where the GW source emission is enhanced by the strength of the DF, but has no corresponding energy loss by the DF mediator emission.
On the other hand, the binary formation becomes significantly more complicated due to the range of the DF. As we show in this work, this can lead to binaries forming predominantly with smaller initial separation than the gravity-only case but larger initial eccentricity than the massless-mediator case, both of which enhance the merger rate. The enhanced merger rate can then lead to enhancements or suppressions in the SGWB, similar to the discussion of (c) in the previous paragraph. Together, these effects can generate distinctive knee, ankle, and plunging features in the SGWB at different GW frequencies, some of which are directly related to the mediator mass and hence provide observational targets and potential confirmations of underlying models.

Models for MDM with a DF were discussed in \cite{Bai:2023lyf} involving both scalar and vector force mediators. Generally, they take the form of a compact object like a dark quark nugget \cite{Bai:2018dxf} (or many other similar models \cite{Rosen:1968mfz,Friedberg:1976me,Coleman:1985ki,Lee:1988ag,Gulamov:2015fya,Brihaye:2015veu,Heeck:2021zvk,Bai:2021mzu,Macpherson:1994wf,Hong:2020est,DelGrosso:2023trq,Wise:2014jva,Wise:2014ola,Hardy:2014mqa,Gresham:2017zqi,Gresham:2017cvl,Kouvaris:2015rea,Giudice:2016zpa,Maselli:2017vfi,Curtin:2019lhm,Curtin:2019ngc,Hippert:2021fch,Gross:2021qgx,Ryan:2022hku,Banks:2023eym,Bai:2023mfi,Bai:2019zcd}), where the quark constituents either have a gauge charge or have a Yukawa coupling to a light scalar field. It is trivial to include a mediator mass in those models. Such MDM could form in the early universe via a first order phase transition or a solitosynthesis/darkleosynthesis-type process \cite{Bai:2018dxf,Krnjaic:2014xza,Bai:2018vik,Liang:2016tqc,Hong:2020est,Lu:2022paj,Kawana:2022lba,Bai:2022kxq}. If they are only a fraction of the total DM density, the rest of DM could be made up of particles that did not become bound in the MDM. Ultimately, this work takes a model-independent parametrization of the underlying MDM model, and while the DF force is taken to be vector mediated, the formulae and results for a scalar-mediated DF only differ by a few factors of two. With the simplifying assumption that all MDM have the same mass and carry the same magnitude of DF charge, the force $F$ between two MDM separated by a distance $d$ can be written as
\begin{equation} \label{eq:force_balance_real}
    F=-\dfrac{G\mdm^2}{d^2}\Big(1+(\GpoG-1)(1+\mmed d)e^{-\mmed d}\Big) \,
\end{equation}
where $\mdm$ is the MDM mass, $\mmed$ is the mass of the DF mediator, $G$ is Newton's constant, and the DF enhancement factor $\GpoG$ expresses the combined strength of the DF and gravity in the limit $d \ll \mmed^{-1}$ as a multiple of $G$. For example, $\beta - 1 = - g^2 q_1 q_2 / (4 \pi G \mdm^2)$ for MDM of charge $q_{1,2}$ with a vector mediator with the gauge coupling $g$, and similarly for a scalar mediator with $-g^2$ replaced by the squared Yukawa coupling $y^2$. 

As an independent probe of DM involving only gravity, weak lensing sets constraints on MDM between asteroid and solar mass to have abundance less than about $\mc{O}(10^{-3})$ of the total DM abundance \cite{Niikura:2017zjd,Smyth:2019whb,Griest:2013aaa,Macho:2000nvd,EROS-2:2006ryy,Wyrzykowski_2011,Niikura:2019kqi,Zumalacarregui:2017qqd,Oguri:2017ock,Mroz:2024wia}, depending on the exact mass and radius \cite{Croon:2020wpr,Bai:2020jfm}, and there are proposals to test still lower masses that are currently unconstrained \cite{Katz:2018zrn,Bai:2018bej,Tamta:2024pow,Jung:2019fcs,Fedderke:2024wpy}.
Above solar mass, there are even stronger constraints \cite{Ricotti:2007au,Ali-Haimoud:2016mbv,Poulin:2017bwe,Serpico:2020ehh,Bai:2020jfm,Brandt:2016aco,Koushiappas:2017chw,Afshordi:2003zb,Murgia:2019duy}. Measurements of the SGWB have the potential to provide complementary and in some cases stronger bounds for MDM in the subsolar-solar mass range.

The remainder of this work is organized as follows. Sec.~\ref{sec:model} provides more details on possible MDM models with a DF. Sec.~\ref{sec:evo} describes the orbital evolution, lifetime, and GW emission for binaries with a variety of possible configurations. In Sec.~\ref{sec:distribution}, the formation of MDM binaries in the earlier universe is discussed and the statistical distribution of MDM binary initial conditions is obtained. Sec.~\ref{sec:SGWB} puts all of the ingredients from the previous sections together to calculate and analyze the SGWB from MDM binaries. Finally, discussion and conclusions are presented in Sec.~\ref{sec:conclusion}. Details about parameter definitions, including a table of frequently used variable names, are given in the Appendices.

\section{Macroscopic dark matter with a finite-range dark force}
\label{sec:model}
For an individual MDM with an approximately constant internal energy density $\rhodm$ and a spherical shape, the relation between its mass $\mdm$ and radius $\rdm$ is 
\beqa
\label{eq:Mmdm}
\mdm = \frac{4\pi}{3}\rhodm\,\rdm^3 = 0.05\,M_\odot \,\left(\frac{\rhodm}{(0.1\,\mbox{GeV})^4} \right)\,\left( \frac{\rdm}{10\,\mbox{km}} \right)^3 ~.
\eeqa
The above relation holds when one can ignore the gravitational energy as well as the self-interaction energy from the long-range force. The former requires that $G\mdm \ll \rdm$ or 
\beqa
\mdm \ll \left( \frac{4\pi}{3}\rhodm\,G^{3}\right)^{-1/2} = 75\,M_\odot\,\left(\frac{\rhodm}{(0.1\,\mbox{GeV})^4} \right)^{-1/2}~. 
\eeqa

For a long-range force mediated by a (Abelian) vector boson with  mass $\mmed \ll 1/\rdm$, an additional screening effect to suppress the static Coulomb charge has to be taken into account. If the dark object is made of many constituents like the dark quark nuggets~\cite{Bai:2018dxf}, the total effective charge under the Abelian vector symmetry is related to the ``Debye length" with $\lambda_{\rm D} = \sqrt{\pi/(2\,\mathfrak{g}\,g^2\,\mu_{\rm eq}\,\sqrt{\mu_{\rm eq}^2 + m^2 })}$~\cite{Jancovici}. Here, $\mathfrak{g}$ represents the degrees of freedom of the degenerate fermion constituents, $\mu_{\rm eq}$ is the equilibrium value
for the chemical potential, and $m$ is the constituent fermion mass. Requiring the Debye length to be longer than the size of the MDM and noting that $\lambda_{\rm D}$ is maximized when $m=0$, there is an upper bound on the fifth force between two MDM with $d \ll 1/\mmed$~\cite{Bai:2023lyf} 
\beqa
\beta - 1  = \frac{\mathfrak{g}^{1/2}\,g^2}{8\sqrt{3}\pi^2\,G\,\rhodm^{1/4}} < 14\times\left( \frac{(0.1\,\mbox{GeV})^4}{\rhodm} \right)^{1/3} \, \left( \frac{10^{-3}\,M_\odot}{\mdm} \right)^{2/3}~. 
\eeqa
For a less-dense and a lighter MDM, one could have a large value of $\beta$. Also note that the self-interaction potential energy from the Abelian gauge force can be ignored using the above findings. 

Other than the nontopological solitons like dark quark nuggets~\cite{Bai:2018dxf} composed of fermionic constituents, other solitons with bosonic constituents may also experience long-range or finite-range forces when the constituents are charged under additional vector-like or scalar-like forces~\cite{Lee:1988ag,Gulamov:2015fya,Brihaye:2015veu,Heeck:2021zvk,Bai:2021mzu}. The formation of those nontopological solitons has been studied in Ref.~\cite{Bai:2022kxq}. Beyond solitons, the binaries of dark-charged primordial black holes in modified gravity scenarios could have an additional finite-range dark force~\cite{Cardoso:2016ryw}.

\section{Binary orbit evolution with a finite mediator mass}
\label{sec:evo}
Consider the general case of a central force between two MDM objects, which guarantees the trajectories of the objects are confined on a plane. 
In the center-of-mass (COM) frame, the radial equation of motion without dissipation for one of the MDMs at a distance $r_1$ from the COM is 
\begin{align}\label{eq:EOM_Newton}
\ddot{r}_1=\dfrac{J^2}{\mdm^2 r^3_1}-\dfrac{1}{2\mdm}\dfrac{d}{dr_1}U(r_1)
\end{align}
where $J=\mdm r^2_1\dot\theta$ is the orbital angular momentum of {\it an individual} MDM, $U(r_1)$ is the potential energy of the binary system, and dots represent derivatives with respect to time.
{\it E.g.}, for the gravity-only case $U(r_1)=-G\mdm^2/(2r_1)$. 
A common practice is to combine the RHS into an effective potential, \ie,
\begin{align}
U_{\rm eff}(r_1)&=\dfrac{J^2}{2\mdm r^2_1}+\dfrac{1}{2}U(r_1)\,,
\end{align}
such that the radial motion of the MDM can then be viewed as a one-dimensional motion under $U_{\rm eff}$ as $\ddot{r}_1=-U^\prime_{\rm eff}/\mdm$.
The period of the radial motion is given by
\begin{align} \label{eq:period_integral}
T=2\int^{r_{1,{\rm max}}}_{r_{1,{\rm min}}}\sqrt{\dfrac{\mdm}{E-U_{\rm eff}(r_1)}}\,,
\end{align}
where $r_{1,{\rm max}}$ and $r_{1,{\rm min}}$ are the maximum and minimum orbital radii of the MDM, obtained by solving
$E=U_{\rm eff}(r_1)$,  with $E$ the total energy of the binary.

\subsection{Binary evolution with gravitational and long-range dark forces}\label{subsec:evo_grav}

\subsubsection{Orbital evolution without DF mediator emission}
\label{subsubsec:evo_grav}
The evolution of binary systems with only gravity has been well-summarized in literature like~\cite{Maggiore:08textbook}.
Here we briefly review several key results which benefit our later discussion of finite-ranged dark forces.  
Results for the gravity-only case are denoted with the subscript `GR'.

With gravity being the only interaction between the binary, the binary system follows the ordinary Kepler's law, where the period of the motion can be written as
\begin{align}\label{eq:Kepler}
T=\left(\dfrac{2\pi^2 a^3}{G\mdm}\right)^{1/2}\,,
\end{align}
with $a$ as the semi-major axis.
The total energy of the system follows the virial theorem
\begin{align}\label{eq:virial}
E=-\dfrac{G\mdm^2}{2a}\,.
\end{align}
The semi-major axis $a$ and the eccentricity $e$ of the elliptic orbit are related to the total energy $E$ and angular momentum of an individual object $J$ via
\begin{align}\label{eq:EJtoae}
a=-\dfrac{G\mdm^2}{2E}\,,\qquad e=\dfrac{\sqrt{G^2\mdm^6+16EJ^2\mdm}}{G\mdm^3}\,.
\end{align}

The energy and angular momentum radiation via GW emission can be calculated from the second and third time derivative of the energy density's traceless quadruple moment $Q_{ij}$ as
\begin{align}\label{eq:EdotJdot_GW}
\dot{E}_\text{GW} = \frac{G}{5} \langle \dddot{Q}_{ij} \dddot{Q}_{ij} \rangle \, ,\qquad
\dot{J}_\text{GW}^i = \frac{G}{5} \epsilon^{ikl} \langle \ddot{Q}_{ka} \dddot{Q}_{la} \rangle \, .
\end{align}
Note that here and afterwards the coefficient of $\dot{J}$ differs from results in some literature (\eg, Ref.~\cite{Maggiore:08textbook}) by a factor of two due to the definition of $J$. 
In practice, it is conventional to take the period-averaged results for the emission rates
\begin{align}\label{eq:dEdt_GR}
\left(\dfrac{dE}{dt}\right)_{\rm GR}&=\dfrac{(\Delta E)_{\rm GR}}{T}
=-\dfrac{(-E)^{3/2} G^2\mdm^{5/2}(9472 E^2J^4+5856 G^2EJ^2\mdm^5+425 G^4\mdm^{10})}{3840J^7}\,,\\
\label{eq:dJdt_GR}
\left(\dfrac{dJ}{dt}\right)_{\rm GR}&=\dfrac{(\Delta J)_{\rm GR}}{T}=-\dfrac{(-E)^{3/2} G^2\mdm^{5/2}(112EJ^2+15G^2\mdm^5)}{20J^4}\,,
\end{align}
where $\Delta E$ and $\Delta J$ are the change of $E$ and $J$ over one period, which in this scenario is caused only by GW emission.
Specifically, the emitted energy is
\begin{align}
(\Delta E)_{\rm GR}=\dfrac{\pi G^3 \mdm^5}{7680J^7}\left(9472E^2J^4+5856 E G^2 J^2 \mdm^5+425 G^4\mdm^{10}\right)\,.
\end{align}

The lifetime of the binary is mathematically easier to estimate with $a$ and $e$, whose evolution $\dot{a}$ and $\dot{e}$ can be derived in terms of $(a, e)$ with the help of Eq.~\eqref{eq:EJtoae}. 
In terms of the initial values $a_0$ and $e_0$ and ignoring the finite radius of the MDM, the merger lifetime using the fact that $e$ monotonically decreases is calculated as
\begin{align}\label{eq:tau_GR}
\tau_{\rm GR}=&\int^0_{e_0}\dfrac{1}{\dot{e}}de =\dfrac{15 a^4_0(1-e^2_0)^{7/2}}{608\uu G^3 \uu \mdm^3 e^2_0}h_{\rm gv}(e_0)\,,\\
h_{\rm gv}(e_0)=&\left(1+\frac{121}{304}e^2_0\right)^{-1} -\sqrt{1-e^2_0}\left(1+\frac{121}{304}e^2_0\right)^{-\frac{3480}{2299}}\,\left[ F_1\left(\frac{5}{19};\frac{1}{2},\frac{1118}{2299};\frac{24}{19};e^2_0,-\frac{121}{304}e^2_0\right)\nonumber\right.\\
&\left.+\frac{47}{192}e^2_0 F_1\left(\frac{24}{19};\frac{1}{2},\frac{1118}{2299};\frac{43}{19};e^2_0,-\frac{121}{304}e^2_0\right) \right]\,,
\end{align}
where $F_1(a;b_1,b_2;c;x,y)$ is the first Appell function~\cite{olver2010nist}.
The function $h_{\rm gv}(e_0)$ is monotonic, with $h_{\rm gv}(e_0 \to 0) \sim \frac{19}{48} e_0^2$ and $h_{\rm gv}(1)=\frac{304}{425}$.
The emitted GW spectrum $dE/d\fGWs$ during the merger is given by the classical $\fGWs^{-1/3}$ power-law as
\begin{align}\label{eq:dEGWdf_GR}
\left(\dfrac{dE_{\rm GW}}{d\fGWs}\right)_{\rm GR}=\left(\dfrac{\pi^2G^2\mdm^5}{54\fGWs}\right)^{1/3}.
\end{align}
The subscript `s' refers to the emission in the binary source frame; redshift effects will be taken into account later. This and all following formulas assume GWs are emitted during any given orbit exclusively by $\fGWs=2/T$, ignoring harmonics with frequencies at integer and half-integer multiples of this, as justified in \cite{Bai:2023lyf}. 

Before moving to the results for a DF with a massless mediator, we address how the results above change when an additional inverse-square attractive force {\it that does not radiate} is imposed on the binary.
This corresponds to the case where $\mmed d \ll 1$ so that the strength of the dark force is active in $F$ of Eq.~\eqref{eq:force_balance_real}, while $T^{-1} \ll \mmed$ so that the emission of its mediator has not yet initiated. 

Naively, one may expect the expressions in this case (denoted by subscript `$\GpoG$GR') can be easily obtained from the GR ones by the simple rescaling $G\to\GpoG G$, which is the case for Eqs.~\eqref{eq:Kepler},~\eqref{eq:virial}, and~\eqref{eq:EJtoae}.
However, the emission of GWs involves the coupling to the graviton, a factor of $G$ that should not be rescaled.
Hence, for the energy emission one finds
\begin{align}
(\Delta E)_{\GpoG{\rm GR}}&=\dfrac{\pi \GpoG^2 G^3 \mdm^5}{7680J^7}\left(9472E^2J^4+5856 E \GpoG^2 G^2 J^2 \mdm^5+425 \GpoG^4 G^4\mdm^{10}\right)\,,\\
\label{eq:dEdt_bGR}
\left(\dfrac{dE}{dt}\right)_{\GpoG{\rm GR}}
&=-\dfrac{(-E)^{3/2} \GpoG G^2\mdm^{5/2}(9472 E^2J^4+5856 \GpoG^2G^2EJ^2\mdm^5+425\GpoG^4G^4\mdm^{10})}{3840J^7}\,.
\end{align}
The expressions for $J$ can be similarly derived.
The lifetime of the binary rescales to
\begin{align} \label{eq:tau_betaGR}
\tau_{\GpoG{\rm GR}}=\dfrac{1}{\GpoG^2}\tau_{\rm GR}\,.
\end{align} 
The emitted GW spectrum $dE_{\rm GW}/d\fGWs$ is enhanced by a factor of $\GpoG^{2/3}$ compared with the gravity-only case:
\begin{align}\label{eq:dEGWdf_betaGR}
\left(\dfrac{dE_{\rm GW}}{d\fGWs}\right)_{\GpoG{\rm GR}}=\left(\dfrac{\pi^2\GpoG^2G^2\mdm^5}{54\fGWs}\right)^{1/3}.
\end{align}

\subsubsection{Orbital evolution with massless dark force emission} 
\label{sec:evo_masslessDF}

When the mediator of the dark force is massless, the DF range is infinite and the dipole radiation of the dark force is always active. 
We denote the expressions where the emission of dark force mediators has initiated with the subscript `DF'.
As shown in~\cite{Bai:2023lyf}, the emission rates can be calculated with the dipole momentum $p_i$ as\footnote{The expressions here assume a massless vector mediator, enabling direct comparison to \cite{Bai:2023lyf}, but expressions for the massless scalar mediator differ only by a factor of two. One subtlety for the scalar mediator is that there is no dipole scalar radiation when the MDM charges are exactly equal because the charges are of the same rather than opposite sign. However, realistically the MDM charges should differ by $\mathcal{O}(1)$ factors, which in turn changes $(dE/dt)_\text{DF}$ and $(dJ/dt)_\text{DF}$ only a little.}
\begin{align}\label{eq:EdotJdot_DF}
\dot{E}_\text{DF} = -\dfrac{1}{6 \pi } \ddot{p}^2 \, ,\quad
\dot{J}^i_\text{DF} = -\dfrac{1}{12\pi}\epsilon^{ijk}\dot{p}_j\ddot{p}_k\,,
\end{align}
which after period-averaging gives
\begin{align}\label{eq:dEdt_DF}
\left( \dfrac{dE}{dt} \right)_{\rm DF} &= \dfrac{4\GpoG^2(\GpoG-1)G^3\mdm^4(2+e^2)}{3a^4(1-e^2)^{5/2}}\,,\\
\label{eq:dJdt_DF}
\left( \dfrac{dJ}{dt} \right)_{\rm DF} &= \dfrac{2\sqrt{2}\GpoG^{3/2}(\GpoG-1)G^{5/2}\mdm^{7/2}}{3a^{5/2}(1-e^2)}\,.
\end{align}
$E$ and $J$ are related to $a$ and $e$ via
\begin{align}\label{eq:EJtoae_DF}
a=-\dfrac{\GpoG G\mdm^2}{2E}\,,\quad e=\dfrac{\sqrt{\GpoG^2G^2\mdm^6+16EJ^2\mdm}}{\GpoG G\mdm^3}\,,
\end{align}
which is Eq.~\eqref{eq:EJtoae} after the replacement $G\to\GpoG G$.
During the evolution, $a$ and $e$ are related as
\begin{align}
\label{eq:aeDF}
a(e) = a_0 \frac{g(e)}{g(e_0)} \, ,
\quad \quad
g(e) = \frac{e^{4/3}}{1-e^2} \, ,
\end{align}
where $a_0$ and $e_0$ are again the initial values of $a$ and $e$. With a strong enough DF, the merger lifetime is
\begin{align}\label{eq:tau_DF}
\tau_{\rm DF}=\dfrac{a^3_0}{4\GpoG(\GpoG-1)G^2\mdm}\dfrac{(1-e^2_0)^{5/2}(1-\sqrt{1-e^2_0})^2}{e^4_0}\,.
\end{align}
Note that the expression above assumes DF radiation dominates GW radiation and hence one should not take $\beta\to 1$ where the DF vanishes. However, this assumption does remain valid even for $\beta - 1 \ll 1$ because dipole emission tends to dominate quadrupole emission (provided $\tau_{\rm DF} < \tau_{\rm GR}$).

The emitted GW spectrum from the binary is
\begin{align} \label{eq:dEdfGWs_DF}
\left(\dfrac{dE_{\rm GW}}{d\fGWs}\right)_{\rm DF}
&=\dfrac{\pi\sqrt{a}  \left(37 e^4+292 e^2+96\right) (\GpoG G)^{3/2} \mdm^{5/2}}{3\sqrt{2}\left[10\uu a(1-e^2)(2+e^2)(\GpoG-1)+\left(37 e^4+292 e^2+96\right)(\GpoG G) \mdm\right]}\,,
\end{align}
where $e$ can be expressed as a function of $a$, $a_0$, and $e_0$ using (\ref{eq:aeDF}), and $a$ further expressed in terms of $\fGWs$ using the rescaled Kepler's law (Eq.~\eqref{eq:Kepler} after replacing $G\to\GpoG G$). Thus, $dE_{\rm GW}/d\fGWs$ can be expressed as a function of only $\fGWs$ and the initial conditions $(a_0,e_0)$ when the binary starts evolving with an effectively massless DF. Compared to that with only gravity~\eqref{eq:dEGWdf_GR}, the source emission spectrum with DF mediator emission is suppressed at low frequency while enhanced at the high frequency (in both the amplitude and the cut-off), as the DF not only accelerates the orbiting but also leads to a quicker evolution.

\subsection{Orbital evolution with a finite-ranged attractive dark force}\label{subsec:evo_DF}
\begin{figure}[t]
    \centering
    \includegraphics[width=0.45\linewidth]{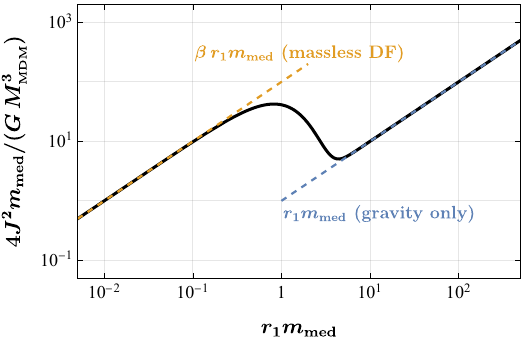}
    \hspace{3mm}
    \includegraphics[width=0.45\linewidth]{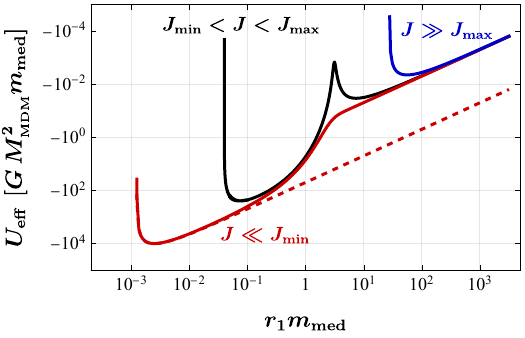}
    \caption{
    {\it Left:} 
    An example of the LHS of~\eqref{eq:dUeff_numerator} when $\GpoG>\GpoGbarrier$. For some values of $J$ there may be either a single extremum or three extrema in the effective potential, corresponding to the red/blue curves and the black curve in the right panel, respectively.
    {\it Right:} Examples of the effective potential in the regimes $J\gg J_{\rm max}$ (blue), $J_{\rm min}<J<J_{\rm max}$ (black), and $J\ll J_{\rm min}$ (red solid) with $J_{\rm max}$ and $J_{\rm min}$ given in \eqref{eq:J_max} and \eqref{eq:J_min}. In the black curve, $\Ubarrier$ sits at $r_{1,{\rm barrier}}\mmed \approx 3$, separating the gravity-like region at larger $r_1$ and the DF region at smaller $r_1$.
    The dashed red curve shows $U_{\rm eff}$ when the DF mediator is massless (plotted with the same finite $\mmed$ as the solid curves for the ``unit'' of the $x$- and $y$-axes).
    }
    \label{fig:dUeff_and_Ueff}
\end{figure}

When the dark force mediator is massive, it is possible for an individual binary to evolve through all three of the stages described in Sec.~\ref{subsec:evo_grav}---GR, $\GpoG$GR, and DF---in addition to hybrid or transition stages not fully encompassed by any of the prior description.
The mass of the dark force mediator, \ie, the range of the dark force, is incorporated into $\Ueff$ as a Yukawa term using the force equation in~\eqref{eq:force_balance_real}.
With the mediator mass denoted as $\mmed$, the effective potential of the binary system is
\begin{align}\label{eq:Ueff}
U_{\rm eff}=\dfrac{1}{2}\,\left(-\dfrac{G\mdm^2}{2r_1}-\dfrac{(\GpoG-1)G \mdm^2 e^{-2\mmed r_1}}{2r_1}\right)+\dfrac{J^2}{2\mdm r^2_1}\,.
\end{align}
Naively, the exponential function seems to suggest a smooth transition between the gravity-only and the massless-DF scenarios reviewed earlier.
However, the real situation can be more subtle, as the Yukawa interaction may actually generate a barrier in the effective potential, shown in the right panel of Fig.~\ref{fig:dUeff_and_Ueff}. On the two sides of the barrier, $\Ueff$ behaves as if there is only gravity and as if the DF is long-ranged. We thus refer to the effective potential around the two local minima as the ``gravity-like region'' and ``DF region'', respectively. 

Examining the extrema of the effective potential by checking $d\Ueff/dr_1=0$, one finds
\begin{align}\label{eq:dUeff_numerator}
\mmed\uu r_1+(\GpoG-1)e^{-2\mmed\uu r_1}\mmed\uu r_1(1+2\mmed\uu r_1)&=\dfrac{4J^2\mmed}{G\mdm^3}\,,
\end{align}
where the $J$- and $r_1$-dependencies are separated to the RHS and LHS of the equation.
The LHS of~\eqref{eq:dUeff_numerator} always has a decreasing knee when $\GpoG > 1+ e^3/5 \equiv \GpoGbarrier$ (with $e$ here indicating Euler's constant rather than eccentricity), as demonstrated in the left panel of Fig.~\ref{fig:dUeff_and_Ueff}. Thus, \eqref{eq:dUeff_numerator} has three solutions when $J$ is within a range $J_{\rm min}<J<J_{\rm max}$ corresponding to two local minima separated by a barrier in $\Ueff$.
With $\GpoG$, $\mdm$, and $\mmed$ fixed, $\Ueff(r_1)$ is uniquely determined by $J$ and we therefore denote the barrier height as $\Ubarrier(J)$ and the location of the barrier as $r_{1,{\rm barrier}}$.
Outside the decreasing knee region, when $J\gg J_{\rm max}$, $\Ueff$ behaves like the gravity-only case with the dark force inactive between the binary.
For $J\ll J_{\rm min}$, on the other hand, $\Ueff$ behaves like the massless-DF case at small orbital radii $\mmed\uu r_1 \ll 1$ and smoothly transits to the gravity-only case as $\mmed\uu r_1\gg 1$.
See the right panel of Fig.~\ref{fig:dUeff_and_Ueff}. Here, one has 
\beqa
\label{eq:J_max}
J_{\rm max} &=& \frac{\sqrt{G}\,\mdm^{3/2}\,\mmed\,r_{1,2}^{3/2} }{\sqrt{4\,\mmed^2\,r_{1,2}^2 -2 \mmed\,r_{1,2}-1}} ~,\\
\label{eq:J_min}
J_{\rm min} &=& \frac{\sqrt{G}\,\mdm^{3/2}\,\mmed\,r_{1,1}^{3/2} }{\sqrt{4\,\mmed^2\,r_{1,1}^2 -2 \mmed\,r_{1,1}-1}} ~,
\eeqa
with $r_{1,2}$ and $r_{1,1}$ as the larger and smaller zeros of the following equation
\beqa
e^{2\,\mmed\,r_1} - (\beta -1)\Big(2\mmed\,r_1(2\mmed\,r_1-1) - 1\Big) = 0 ~. 
\eeqa
For the larger zero with $\mmed r_{1,2} \gg 1$, one has $\mmed r_{1,2}\approx -W_{-1}(-\sqrt{1/(\beta-1)}/2)$ where $W_{-1}(x)$ with $-1/e \leqslant x < 0$ is the Lambert function in the second branch. In the limit of $\beta \rightarrow +\infty$ or $x \rightarrow 0^-$, one has $-W_{-1}(x) \approx -[\ln(-x) - \ln(-\ln(-x))]$~\cite{2020arXiv200806122L}. There is a simpler approximate formula for $J_{\rm max} \approx \sqrt{G}\,\mdm^{3/2}\,r^{1/2}_{1,2}/2$.

Note that $J$ is not a fixed value but decreases with time due to GW and DF mediator emission.
The evolution of a MDM binary acting under the influence of a massive DF mediator is thus more complicated than that in the gravity-only and massless-DF cases, as the configuration of $\Ueff$ changes between the three categories shown in the right panel of Fig.~\ref{fig:dUeff_and_Ueff} during the course of evolution.
Binary systems with initial angular momentum $J_0 > J_\text{max}$ (the blue curve in the right panel of Fig.~\ref{fig:dUeff_and_Ueff}) may have the barrier in $\Ueff$ develop later on, and any barrier in $\Ueff$ will eventually disappear later in the binary evolution.
It is then particularly important to track whether $\Ueff$ has a barrier and, if yes, its corresponding position, because the orbit of the binary system may change drastically before and after the point of $E=\Ubarrier$.

An additional complication comes from the emission of the dark force mediator.
Compared to the massless-DF case where it is always active, the DF mediator emission can be triggered only after passing the threshold $\fGWs>\mmed$~\cite{Chen:2024ery}.
For concreteness, we assume that a binary experiences the force of the new mediator before it begins emitting the new mediator radiation.\footnote{If DF emission turns on before the dark attractive force, the GW spectrum is always suppressed. Thus, we do not consider such cases.} In other words, $\mmed^{-1} > (2 \GpoG G \mdm / (2 \pi \mmed)^2)^{1/3} \equiv \aemit$, where $\aemit$ comes from Kepler's law with $\fGWs=\mmed$. This is valid for $\mmed < \frac{2 \pi^2}{G \mdm} \approx 3 \times 10^{-5}~\text{eV} \left(\frac{10^{-4} M_\odot}{\mdm} \right)$, where $G$ is used in place of $G'$ in the inequality because this tests whether mediator emission turns on before the DF is within its effective range.

\begin{figure}[t!]
    \centering
    \includegraphics[width=0.45\linewidth]{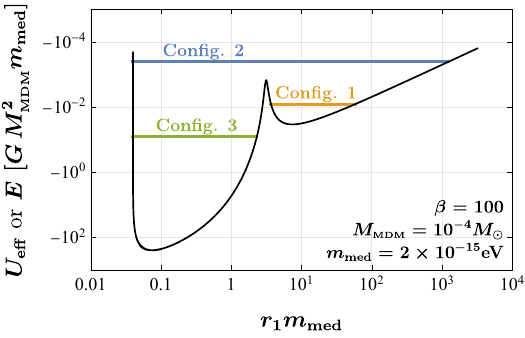}
    \includegraphics[width=0.45\linewidth]{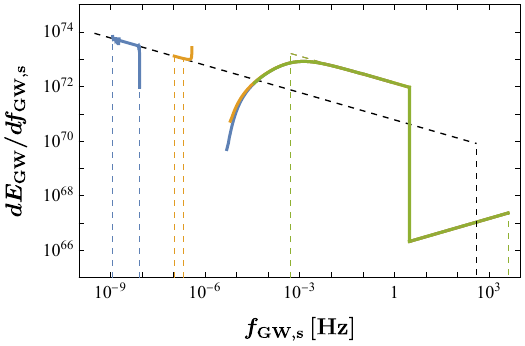}
    \caption{
    {\it Left:} The three benchmark binary initial configurations (horizontal lines) whose evolutions are examined numerically, corresponding to the first three configurations described in Sec.~\ref{subsec:evo_DF}. All benchmarks share the same $J_0=2.93\times 10^{72}$.
    The initial energy $E_0$ of the benchmarks are
    $E_0/(-G\mdm^2\mmed)=0.0004$, 0.008, and 0.08 for the blue, orange, and green lines, respectively.
    {\it Right:} The numerically calculated (solid) and analytically approximated (Eq.~\eqref{eq:dEdfGWs_analytic}, colored dashed) $dE_{\rm GW}/d\fGWs$ for the three benchmark configurations, matched to the left panel by color. 
    The dashed black curve is the binary GW emission spectrum of the gravity-only case.
    The high-frequency cutoff to the spectra depend on the internal MDM density, which here is fixed to $\rhodm=(0.1~\text{GeV})^4$.
    }
    \label{fig:dEGWdf_numeric}
\end{figure}

To understand the orbital evolution and the emitted GW spectrum, we numerically examine the evolution of a binary in terms of $E$ and $J$, which uniquely describe the effective potential and the binary orbit (up to a bifurcation ambiguity, see later discussions in Secs.~\ref{sec:config2} and~\ref{sec:config3}).
During the evolution, the changing rates $dE/dt$ and $dJ/dt$ are calculated fully numerically by averaging Eqs.~\eqref{eq:EdotJdot_GW} and~\eqref{eq:EdotJdot_DF} [with \eqref{eq:EdotJdot_DF} only included when $\fGWs>\mmed$] over one period of the numerically solved radial orbital motion. The emitted GW spectrum is calculated as
\begin{align}\label{eq:dEdf_calc}
\dfrac{dE_{\rm GW}}{d\fGWs}=\dfrac{dE_{\rm GW}/dt}{d\fGWs/dt}\,.
\end{align}
We present the numerical results of three benchmark cases and their corresponding emitted GW spectra in Fig.~\ref{fig:dEGWdf_numeric}. All benchmarks share the same $J_0$ and $\GpoG$ such that they have the same $\Ueff$ with a barrier at the beginning of their evolution. We refer to these benchmarks in the following analysis.
 
\begin{figure}[t!]
    \centering
    \includegraphics[width=0.9\linewidth]{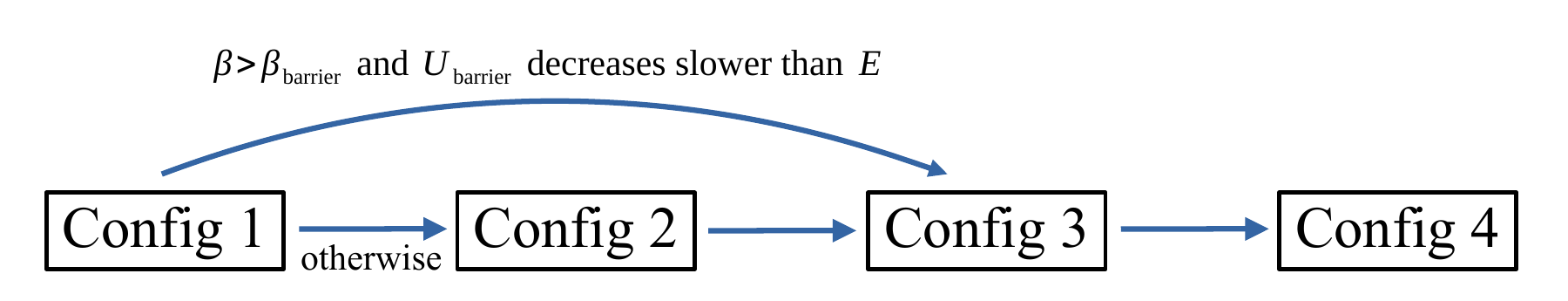}
    \caption{A schematic of the time-ordering of the possible configurations a binary can evolve through. Configuration definitions and numbering are as in Sec.~\ref{subsec:evo_DF}; see text for a full description of the criteria to transition between each configuration.
    }
    \label{fig:config_schematic}
\end{figure}

In the remainder of this subsection, we break down the orbit into four possible configurations according to the relationship between $\Ueff$ and $E$. In general, a binary will evolve through several of these configurations throughout its inspiral, and thus its GW emission will be characterized by pieces from each configuration. 
Configurations 1, 3, and 4 are all familiar from the prior discussion, corresponding approximately to the GR, $\GpoG$GR, and DF cases, respectively. Configuration 2 is unique to the case of a massive DF mediator. A schematic diagram of how binaries evolve through each of these configurations is given in Fig.~\ref{fig:config_schematic}. The starting configuration depends on a binary's initial conditions.
For the examples in Fig.~\ref{fig:dEGWdf_numeric}, the system indicated by the orange lines evolves through Configurations 1, 3, and 4 in sequence; the blue system evolves through 2, 3, and 4; and the green system evolves through only 3 and 4 (a system starting in Configuration 4 would look similar to the green line, just with lower $E$ and $J$ in the left panel and a higher minimum $\fGWs$ in the right panel). Regardless of the evolutionary sequence, a binary spends the majority of its lifetime at the beginning stage of its evolution, so the merger lifetime can generally be well-approximated using only the lifetime formula corresponding to its initial configuration.

\subsubsection{Configuration 1: Binary in a gravity-only region}
\label{sec:config1}

Configuration 1 corresponds to binaries with $a >2\,r_{1,\text{min}} > \mmed^{-1}> \aemit$. If $\GpoG > \GpoGbarrier$, this can occur because either $J>J_{\rm max}$ (as in the blue curve in the right panel of Fig.~\ref{fig:dUeff_and_Ueff}), or $J_{\rm min}<J<J_{\rm max}$ with $E<\Ubarrier$ and the binary to the right of the barrier (as in the orange curves in Fig.~\ref{fig:dEGWdf_numeric}). If $\GpoG < \GpoGbarrier$, this occurs with sufficiently large $J$. 
Binaries in this configuration cannot resolve the DF region of $\Ueff$, and thus their evolution is described by the GR formalism introduced in Sec.~\ref{subsec:evo_grav} (\eg, Eqs.~\eqref{eq:dEdt_GR} and~\eqref{eq:dJdt_GR}).
The GW spectrum emitted by the binary is given by Eq.~\eqref{eq:dEGWdf_GR}, corresponding to the lowest-frequency portion of the orange curve in the right panel of Fig.~\ref{fig:dEGWdf_numeric}.

If a binary starts in such a configuration as its initial condition, its lifetime is dominated by $\tau = \tau_\text{GR}$ as defined in Eq.~\eqref{eq:tau_GR}. In general, we assume that no binary ever evolves from another of the configurations into this configuration. We justify this in the discussion of Configuration 3 below.

\subsubsection{Configuration 2: Binary passing through both the gravity-only and DF regions in a single orbit}
\label{sec:config2}

Configuration 2 corresponds to binaries with $2\,r_{1,\text{max}}  > \mmed^{-1} >2\,r_{1,\text{min}}> \aemit$. If $\GpoG > \GpoGbarrier$, this can occur because either a) $J<J_{\rm min}$ (as in the red curve in the right panel Fig.~\ref{fig:dUeff_and_Ueff}) with $E \gtrsim U(r_1=\mmed^{-1}/2) \equiv \Ucross$,  or b) $J_{\rm min}<J<J_{\rm max}$ with $E>\Ubarrier$ (as in the initial conditions for the blue curves in Fig.~\ref{fig:dEGWdf_numeric}). If $\GpoG < \GpoGbarrier$, the condition is similar to (a).

In this case, $r_1$ traverses both the gravity-like region and the DF region of $\Ueff$ during one period, but DF mediator radiation does not occur. While unlike the other configurations, the orbit does not form closed elliptical patterns, its radial motion is still periodic.
Generally, the majority of the period is spent in the gravity-like region, while the majority of the dissipation occurs in the DF region.
In other words, the orbiting period is approximately given by Eq.~\eqref{eq:Kepler}, which in terms of $E$ is
\begin{align}
T\approx T_{\rm GR}(E)=\dfrac{\pi G}{2}\sqrt{-\dfrac{\mdm^5}{E^3}}\,,
\end{align}
and the emission rate is approximately
\begin{align}\label{eq:dEdt_betaGR}
\dfrac{dE}{dt}&\approx\dfrac{\Delta E_{\GpoG{\rm GR}}}{T_{\rm GR}}\nonumber\\
&=-\dfrac{(-E)^{3/2}\GpoG^2 G^2\mdm^{5/2}(9472 E^2J^4+5856\GpoG^2G^2EJ^2\mdm^5+425\GpoG^4G^4\mdm^{10})}{3840J^7}\,,\\
\label{eq:dJdt_betaGR}
\dfrac{dJ}{dt}&\approx\dfrac{\Delta J_{\GpoG{\rm GR}}}{T_{\rm GR}}=-\dfrac{(-E)^{3/2}\GpoG^2 G^2\mdm^{5/2}(112EJ^2+15\GpoG^2G^2\mdm^5)}{20J^4}\,,
\end{align}
where we explicitly assume that the DF mediator radiation has not yet been triggered. The evolution of the system can then be tracked numerically by expressing $E$ as a function of $J$ (and {\it vice versa}) via
\begin{align}\label{eq:dEdJ_betaGR}
\dfrac{dE}{dJ}=\dfrac{dE/dt}{dJ/dt}=\dfrac{\Delta E/T}{\Delta J/T}=\dfrac{\Delta E}{\Delta J}=\dfrac{9472 E^2J^4+5856\GpoG^2G^2EJ^2\mdm^5+425\GpoG^4G^4\mdm^{10}}{192J^3}\,.
\end{align}

In this case, the spectrum of the emitted GWs is calculated as
\begin{align}\label{eq:dEdfGWs_trans}
\dfrac{dE_{\rm GW}}{d\fGWs}=\dfrac{dE/dt}{d\fGWs/dt}\approx\dfrac{dE/dt}{d(2/T_{\rm GR})/dt}=\dfrac{dE/dt}{\frac{6\sqrt{-E}}{G \mdm^{5/2}}(dE/dt)}\approx \left(\dfrac{\pi^2 G^2\mdm^5}{54\fGWs}\right)^{1/3}\,.
\end{align}
As a result, the spectrum of the emitted GW is the same as that in the gravity-only (GR) case, even though the majority of the GW emission is not produced in the gravity-only region of the potential. This is confirmed by the numerical results for the low-frequency portion of the blue curve in the right panel of Fig.~\ref{fig:dEGWdf_numeric}. Nevertheless, the time spent in this stage is significantly shorter than in the gravity-only case because $dE/dt$ is enhanced.

If a binary starts in such a configuration as its initial condition, the lifetime can be estimated as follows.  Note that in both Eqs.~\eqref{eq:dEdt_GR} and~\eqref{eq:dEdt_betaGR}, the last term in the bracket of each numerator should dominate each expression when using these initial conditions\footnote{One simple (though unrigorous) way to see this is: The large hierarchy between $a_0$ and $\rmin$ implies $e_0 \to 1$. When written in terms of $a$ and $e$, $\Delta E\propto (1+\frac{73}{24}e^2+\frac{37}{96}e^4) \xrightarrow[]{e\to 1} \frac{425}{96}$. This directly matches the coefficients in the last terms in the numerators of Eqs.~\eqref{eq:dEdt_GR} and~\eqref{eq:dEdt_betaGR}, suggesting these terms are the leading contributions.},
which suggests $(dE/dt)_{\GpoG{\rm GR}}\approx\GpoG^6(dE/dt)_{\rm GR}$. Correspondingly, we expect for this category,
\begin{align} \label{eq:tau_mixedrange}
    \taumix= \frac{1}{\GpoG^6} \tau_{\rm GR}(a\approx a_{\rm GR}, e \approx e_{\rm GR}) \, ,
\end{align}
where $a_{\rm GR}$ and $e_{\rm GR}$ are the orbital parameters that approximately describe the portion of the orbit with $r\gtrsim\mmed^{-1}$, \ie, the part or the orbit experiencing only gravity. For more details on the definition of these parameters, see App.~\ref{sec:params_ae}. For the benchmark numeric evaluation of the blue curve in Fig.~\ref{fig:dEGWdf_numeric}, we found this estimation to differ from the numerically determined merger lifetime by only an $\mathcal{O}(1)$ factor.

Otherwise, only some binaries that start in Configuration 1 can evolve to Configuration 2. If $\GpoG < \GpoGbarrier$, this occurs when $J$ is small enough such that $r_{1,{\rm min}} \lesssim \mmed^{-1}/2$. If $\GpoG > \GpoGbarrier$, this may occur in principle when $J_{\rm min}<J<J_{\rm max}$ and $E=\Ubarrier$, but for this to be true $\Ubarrier$ must continue to decrease faster than $E$ after the two are equal. This can be determined numerically by evolving $E$ and $J$ until $J=J_\text{min}$: if at this point $E > \Ucross$, then the binary transitioned from Configuration 1 to 2; otherwise, it transitioned straight from Configuration 1 to 3. More details on this can be found in the following subsubsection. 

\subsubsection{Configuration 3: Binary in the DF dominated region without DF mediator emission}
\label{sec:config3}

Configuration 3 corresponds to binaries with $\mmed^{-1} > a > \aemit$. If $\GpoG > \GpoGbarrier$, this can occur because either a) $J<J_{\rm min}$ (as in the red curve in the right panel of Fig.~\ref{fig:dUeff_and_Ueff}) with $E \lesssim \Ucross$, or b) $J_{\rm min}<J<J_{\rm max}$ with the binary to the left of the barrier (as in the initial conditions for the green curves in Fig.~\ref{fig:dEGWdf_numeric}). If $\GpoG < \GpoGbarrier$, the condition is similar to (a). 

First, consider the evolution within Configuration 3.
Initially, if  $\rmax \sim \mmed^{-1}$, $T$ (and thus $dE/dt$ or $dJ/dt$) cannot be calculated (semi-)analytically. 
However, the evolution of $E$ and $J$ still largely follows the $dE/dJ$ derived in Eq.~\eqref{eq:dEdJ_betaGR} (\ie, from Configuration 2).
This is because the majority of the dissipation still happens in the innermost region of the orbit and hence $\Delta E$ and $\Delta J$ still take the form of $\Delta E_{\GpoG{\rm GW}}$ and $\Delta J_{\GpoG{\rm GW}}$. 
The period, on the other hand, cancels in obtaining $dE/dJ$.
In other words, the evolution of $E$ and $J$ of the system can be smoothly and continuously tracked with the function $dE/dJ$ across the two stages, despite the irregularity and drastic change in the orbital configuration and period. Once $\rmax \ll \mmed^{-1}$, the orbital evolution is very well approximated by the $\beta$GR expressions in Sec.~\ref{subsubsec:evo_grav}. Thus, if a binary starts in such a configuration as its initial condition, its lifetime is dominated by $\tau = \tau_{\GpoG{\rm GR}}$ in Eq.~\eqref{eq:tau_betaGR}.

Binaries evolving through Configurations~1 and~2 may evolve to Configuration~3 when either $i$) $J<J_{\rm min}$ (or a similar condition for $\GpoG < \GpoGbarrier$) with $E \approx \Ucross$, or $ii$) $J_{\rm min}<J<J_{\rm max}$ and $E=\Ubarrier$. For binaries in Configuration 1, it was already explained in Sec.~\ref{sec:config2} that they may evolve directly to Configuration 3 if both $\GpoG > \GpoGbarrier$ and $E$ decreases faster than $\Ubarrier(J)$, or otherwise evolve to Configuration 2 (see Fig.~\ref{fig:config_schematic}).
Which evolutionary path a Configuration 1 binary takes can be determined numerically in the following way. Since the evolutions of $E$ and $J$ for Configuration 2 and 3 follow the same $dE/dJ$, the binary can be evolved numerically starting from ($ii$) without knowing which configuration the binary has evolved to. Once $J=J_\text{min}$, it is trivial to compare $E$ and $\Ucross$ to determine if the binary was in Configuration 2 or 3 starting from ($ii$).\footnote{In other words, we are assuming if $E$ decreases slower than $\Ubarrier$ on exiting Configuration 1, it will always stay above $\Ubarrier$. This follows the same spirit as the earlier assumption that once a system accesses the DF side of the barrier, it will never jump up the barrier.}
Binaries starting in Configuration 2, on the other hand, exclusively evolve to Configuration 3. It may seem that under condition ($ii$), a binary could have its trajectory continue on the $r_1>r_{\rm barrier}$ side of $\Ueff$ and evolve to Configuration 1 instead of 3. However, that the vast majority of $\Delta E$ happens at radii smaller than the barrier position favors evolving to Configuration 3 instead of 1.

When transitioning from Configuration 1 to 2, there is no change to the low frequency portion of the GW source spectrum $dE_{\rm GW}/d\fGWs$ since both configurations follow~(\ref{eq:Kepler}) and~\eqref{eq:dEGWdf_GR}. However, the transition from either of those configurations to Configuration 3 leads to a sudden change of the binary orbital period $T$ as well as a change in the shape of $dE_{\rm GW}/d\fGWs$. The frequency where~\eqref{eq:dEGWdf_GR} ends (denoted as $f_{\rm trans}$) relies on the position of the transition and thus should be determined numerically. For condition ($ii$) in the previous paragraph, the change in $T$ is discontinuous as displayed in the right panel of Fig.~\ref{fig:dEGWdf_numeric}. This is because crossing the barrier forecloses the ability of the binary to traverse the gravity-like region with $r_1 \mmed >1$, which was the main contribution to $T$ prior to crossing the barrier.\footnote{The gap in $dE_\text{GW}/d\fGWs$ being exactly zero rather than simply suppressed is in part a result of the approximation used here that $\fGWs=2/T$, which neglects higher harmonics that could fill the gap during the first stage of the evolution. The gap is also not present if the barrier disappears before the binary can cross it. In any case, the GW emission is still strongly suppressed in the gap, justifying the exactly zero approximation used in~(\ref{eq:dEdfGWs_analytic}).} At the higher-frequency end of the gap, the binary GW emission spectrum $dE_{\rm GW}/d\fGWs$ starts with a gradual rising feature corresponding to orbits with $r_{1,{\rm max}} \sim \mmed^{-1}$. The power-law spectrum at slightly higher frequencies is well approximated by Eq.~\eqref{eq:dEGWdf_betaGR}.
That the strength of the dark force is activated around a distance $\sim\mmed^{-1}$ suggests the power-law starts at a frequency of $\fGWs=f_{\GpoG}\equiv\sqrt{2\GpoG G \mdm\mmed^3}/\pi$.

\subsubsection{Configuration 4: Binary in the DF dominated region with DF mediator emission}
\label{sec:config4}

Configuration 4 corresponds to binaries with $\mmed^{-1} > \aemit > a$. The conditions on $E$ and $J$ are the same as the prior configuration, just with the more stringent condition on $a$ (and thus $E$). While no binary is plotted with this as an initial condition in Fig.~\ref{fig:dEGWdf_numeric}, all the binaries evolve through this stage in the highest-frequency portion of their spectra. The evolution of the system and the emitted GW spectrum in this stage follows the DF expressions in Eqs.~\eqref{eq:aeDF} and~\eqref{eq:dEdfGWs_DF}. This stage ends when the two MDM objects touch, generally when $a= 2 \rdm$, corresponding to a maximum GW emission frequency $\fmaxs =\sqrt{|F(d=2\rdm)|/(\pi^2\mdm\rdm)} \approx \sqrt{\GpoG G \mdm / (4 \pi^2 \rdm^3)}$. Further emission could occur during merger/ringdown-type stages, but this emission is MDM-model dependent and we conservatively neglect it. Note that if $\mmed$ is too large or the internal density $\rhodm \propto \mdm / \rdm^3 \propto \fmaxs^2$ is too small, binaries may never achieve later configurations like this one and instead merge in an earlier configuration.
 
Generally, binaries in Configuration 3 evolve into Configuration 4 when the emission of the DF mediator becomes active, and hence $a_0=\aemit$ should be taken in Eqs.~\eqref{eq:aeDF} and~\eqref{eq:dEdfGWs_DF} if a binary is not formed initially under Configuration 4.
The initial orbital eccentricity $e_0$ therein, on the other hand, has to be determined numerically from the binary's earlier evolution.
If a binary starts in Configuration 4 as its initial condition, its lifetime is given by $\tau = \tau_\text{DF}$ in Eq.~\eqref{eq:tau_DF}. 

\subsection{Approximated GW emission spectrum}
\label{subsec:GW_emi_spec_approx}

We have seen in the previous subsection how the GW spectrum for a given binary depends on its initial conditions. Sometimes, it is possible to obtain very good analytic estimates for the GW spectrum, but for some initial conditions there can exist a range of frequencies for which numerical computations are required. However, calculating the SGWB requires integrating over all possible binary initial conditions. This makes such numerical computations impractical, because each initial condition requires a separate time-consuming computation.

Instead, we employ the following semi-analytic approximation for any given binary's GW spectrum:
\begin{align}\label{eq:dEdfGWs_analytic}
\dfrac{dE_{\rm GW}}{d\fGWs}=\begin{cases}
\left(\dfrac{\pi^2 G^2\mdm^5}{54\fGWs}\right)^{1/3} & f_{0,s}<\fGWs<f_{\rm trans}\vspace{1ex}\\
\left(\dfrac{\pi^2 \GpoG^2 G^2\mdm^5}{54\fGWs}\right)^{1/3} & f_{\GpoG}<\fGWs<\mmed\vspace{1ex}\\
\left(\dfrac{dE_{\rm GW}}{d\fGWs}\right)_{\rm DF} & \mmed<\fGWs<f_{\rm max,s}
\end{cases}\,.
\end{align}
For all other frequencies, the emission is approximately zero, including for $f_{\rm trans} < \fGWs < f_{\GpoG}$.
In this piece-wise spectrum, $f_{0,s}$ is the lowest GW frequency that the binary can emit, corresponding to the emission frequency at the beginning of the evolution as determined by (\ref{eq:period_integral}) and $\fGWs=2/T$. When $f_{0,s}$ is larger than $f_{\GpoG}$ or $\mmed$, the spectrum starts at $f_{0,s}$ in the corresponding part of the spectrum. The quantity $f_{\rm trans}$ is the frequency just before the binary evolves to Configuration 3 as discussed in Sec.~\ref{sec:config3}, obtained numerically from the binary evolution as $f_{\rm trans}=4(-E)^{3/2}/(\pi G \mdm^{5/2})$ using Eqs.~\eqref{eq:virial} and~\eqref{eq:EJtoae}; $f_{\GpoG}$ and $\fmaxs$ are defined in Secs.~\ref{sec:config3} and~\ref{sec:config4}, respectively. The expression for $\left(dE_{\rm GW}/d\fGWs\right)_{\rm DF}$ is given by Eq.~\eqref{eq:dEdfGWs_DF}. Note that there is a gap between $f_\text{trans}$ and $f_{\GpoG}$, which is largely explained by the sudden change of the binary's orbital configuration when $\rmax$ decreases below $\mmed^{-1}$.

This approximation is based on the analytic results obtained in the prior sections, which capture the most important elements of the numerical results as shown in Fig.~\ref{fig:dEGWdf_numeric}. While it does not fully capture the spectrum between $\fGWs \approx 10^{-5}$ and $10^{-3}~\text{Hz}$ in these examples, this region has $dE_\text{GW}/d\fGWs$ suppressed by several orders of magnitude compared to the neighboring parts of the spectrum, and thus should not contribute too greatly to the SGWB.
Note that both Eq.~\eqref{eq:dEdfGWs_analytic} and the numerical spectra plotted in Fig.~\ref{fig:dEGWdf_numeric} conservatively ignore higher harmonics and emission during merger/ringdown.

For the purposes of SGWB detectability, there are a couple of important changes in $dE_\text{GW}/d\fGWs$ compared to the cases with a purely massless or infinitely massive DF mediator. First, $\fmaxs$ is enhanced by a factor of $\sqrt{\GpoG}$ compared to the gravity-only case, allowing the SGWB to extend to higher frequencies. Second, the middle frequency region in (\ref{eq:dEdfGWs_analytic}) is enhanced compared to both other cases. These factors contribute to enhanced detectability for some frequency ranges of the SGWB.

\section{Binary distribution with a finite mediator mass}
\label{sec:distribution}

With the orbital evolution and emitted GW spectrum of individual binaries, the SGWB can be obtained via a convolution with the merger rate of the binaries, as has been studied in the merger of binary primordial black holes or other macroscopic compact objects~\cite{Wang:2016ana,Sasaki:2018dmp,Raidal:2017mfl,Atal:2022zux,Braglia:2022icu,Banerjee:2023brn,Bai:2023lyf,vanDie:2024htf}.
The binary merger rate $R$ is related to the initial distribution $P$ via
\begin{equation} \label{eq:rate}
    R(\{\mathcal{X}\}) = c_\text{charge} \frac{n_\text{MDM}}{2} P(\{\mathcal{X}\}) = c_\text{charge} \frac{3 H_0^2}{8 \pi G} \frac{\fdm \Omega_\text{DM}}{2 \mdm} P(\{\mathcal{X}\}) \, ,
\end{equation}
where $c_\text{charge} = 1~(1/2)$ for a scalar (vector) DF mediator to account for whether the MDM charge carries a sign, $n_\text{MDM}$ is the comoving number density of MDM, $\fdm$ is the fraction of DM made up of MDM, $\Omega_\text{DM}=0.25$ is the abundance of DM as a fraction of the critical density \cite{Planck:2018vyg}, $H_0 = 100 h~\text{km/s/Mpc}$ with $h=0.678$ is the value of the Hubble parameter today \cite{Planck:2018vyg}, and $\{\mathcal{X}\}$ indicates the set of variables $P$ relies on.
Here we follow the approach in Refs.~\cite{Nakamura:1997sm,Ioka:1998nz,Sasaki:2016jop,Bai:2023lyf} (see also \cite{Ali-Haimoud:2017rtz,Raidal:2017mfl,Raidal:2018bbj,Vaskonen:2019jpv,Hutsi:2020sol,vanDie:2024htf}). Assuming a random homogeneous and isotropic spatial distribution for MDMs after their formation, $P$ takes the form of a Poisson distribution in terms of the initial comoving separation $x$ between a nearest-neighbor MDM pair and the initial comoving separation $y>x$ of the next-nearest MDM object to the nearest-neighbor pair's COM, given by 
\begin{align}\label{eq:Pxy}
P(x,y) dx dy  = \frac{9 x^2 y^2}{\bar{x}^6} e^{-(y/\bar{x})^3} dx dy \, ,
\end{align}
where
\begin{align}
\label{eq:xbar}
\bar{x} = \left(\frac{\mdm}{\rho(z_\text{eq})}\right)^{1/3} = \frac{\fdm^{-1/3}}{(1+z_\text{eq})} \left(\frac{8 \pi G \mdm}{3 H_0^2\,\Omega_\text{DM}}\right)^{1/3} \approx 0.1\uu{\rm pc}\left(\dfrac{\mdm}{M_\odot}\right)^{1/3}\left(\dfrac{1}{\fdm}\right)^{1/3} \, ,
\end{align}
is the average separation between MDMs at matter-radiation equality. In natural units, $(0.1\uu {\rm pc})^{-1} \approx 6 \times 10^{-23} \uu {\rm eV}$.
Note that we normalize the scale factor of the FRW metric $R$ to unity at matter-radiation equality (\ie, $R_{\rm eq}\equiv 1$) such that the ratios $x/\bar{x}$ and $y/\bar{x}$ in~\eqref{eq:Pxy} are meaningful.
For the cases of gravity-only and massless DF, $x$ and $y$ can be translated to quantities relevant for the binary orbital evolution straightforwardly~\cite{Nakamura:1997sm,Ioka:1998nz,Bai:2023lyf}, while there will be some subtleties when the DF mediator has a finite mass, which are the targets of this section.

It is assumed in this treatment that MDM forms well before matter-radiation equality. MDM binaries form when the acceleration from the attractive force between any given nearest-neighbor MDM pair is large enough to overcome the velocity of Hubble expansion. Specifically, for a nearest-neighbor MDM pair with physical separation $d=xR$, decoupling occurs when $(\text{acceleration}) \times (\text{Hubble time}) \sim (\text{Hubble expansion velocity})$, or
\begin{equation}
\label{eq:decoupling}
\frac{F}{\mdm} H^{-1} \sim H d \, .
\end{equation}
Decoupling must occur before matter-radiation equality---afterwards, both sides of this equation have the same scale-factor dependence, so no further decoupling occurs (though see, \eg, \cite{vanDie:2024htf} for further refinements in the gravity-only case).
For the gravity-only and massless DF scenario, this implies $\bar{\rho}_\MDM=\rho_r$ and $\bar{\rho}_\MDM=\rho_r/\GpoG$, respectively, where $\bar{\rho}_\MDM\equiv \fdm (\rho_{\rm eq}/2) (\bar{x}^3/(x^3 R^3))$ and $\rho_r = (\rho_\text{eq}/2)(R^4_{\rm eq}/R^4)$ is the radiation energy density of the universe.
That the dark force mediator has a finite mass effectively corresponds to an interpolation between the two scenarios, and we may use $\mmed$ as a boundary and approximately treat the decoupling condition as
\begin{align}\label{eq:Rdec}
\dfrac{\bar{\rho}}{\rho_r}=\begin{cases}
    1 & \text{for  }x\Rdec > \mmed^{-1}\vspace{2mm}\\
    \dfrac{1}{\GpoG} & \text{for  } x\Rdec<\mmed^{-1}
\end{cases}
\Longrightarrow
\dfrac{\Rdec}{R_{\rm eq}}=\begin{cases}
    \dfrac{1}{\fdm}\left(\dfrac{x}{\bar{x}}\right)^3 R^3_{\rm eq} &  \text{for  }x \Rdec > \mmed^{-1}\vspace{2mm}\\
    \dfrac{1}{\GpoG \fdm}\left(\dfrac{x}{\bar{x}}\right)^3 R^3_{\rm eq} & \text{for  }x \Rdec < \mmed^{-1}
\end{cases}\, .
\end{align}

Once a nearest-neighbor pair of MDM decouples from the Hubble flow, it will form a binary whose initial semimajor axis just after decoupling is well approximated by $a_0 \sim x\Rdec$. From (\ref{eq:Rdec}), decoupling can occur in two possible ranges of $x$:
\begin{align}\label{eq:x_range}
\begin{cases}
    x^4>\dfrac{\fdm}{\mmed}\dfrac{\bar{x}^3}{R^4_{\rm eq}} & \text{for  }x \Rdec > \mmed^{-1}\vspace{2mm}\\
    x^4<\dfrac{\GpoG\fdm}{\mmed}\dfrac{\bar{x}^3}{R^4_{\rm eq}} & \text{for  }x \Rdec < \mmed^{-1}
\end{cases}\,,
\end{align}
where the two $x$-ranges overlap.
Physically, the two ranges result from the fact that nearest-neighbor MDM pairs have two ``chances'' to decouple before matter-radiation equality---once when in range and once when out of range of the DF. If the pair is born too far apart or if they fail to decouple when they are close enough together to feel the attractive DF, the pair may instead decouple a while later when they are farther apart and only feel the force of gravity but Hubble expansion has slowed. Thus, where the two ranges overlap in $x$, the bottom line with $x\Rdec<\mmed^{-1}$ should be used because the binary will take the first opportunity to decouple . Then, \begin{align}\label{eq:a0_DF}
a_0=c_1\times \begin{cases}
    \dfrac{1}{\fdm}\dfrac{x^4}{\bar{x}^3}R^4_{\rm eq} & x>l_{\rm th}\vspace{2mm}\\
    \dfrac{1}{\GpoG \fdm}\dfrac{x^4}{\bar{x}^3}R^4_{\rm eq} & x<l_{\rm th}
\end{cases}\,,
\end{align}
where $l_{\rm th}=\left((\GpoG \fdm/\mmed)(\bar{x}^3/R^4_{\rm eq})\right)^{1/4}$ corresponds to the boundary derived from the lower case of Eq.~\eqref{eq:x_range}. The $\mathcal{O}(1)$ numerical factor $c_1$ must be determined by numerical simulation, and it need not be constant with $x$. No numerical simulation has been performed including a DF, so we adopt $c_1 = 0.4$ from numerical simulations of the gravity-only case \cite{Ioka:1998nz}. 

\begin{figure}[t]
    \centering
    \includegraphics[width=0.325\linewidth]{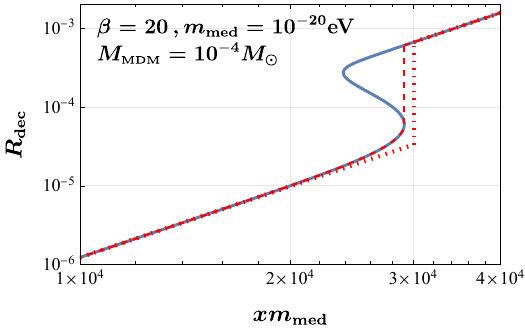}
    \includegraphics[width=0.325\linewidth]{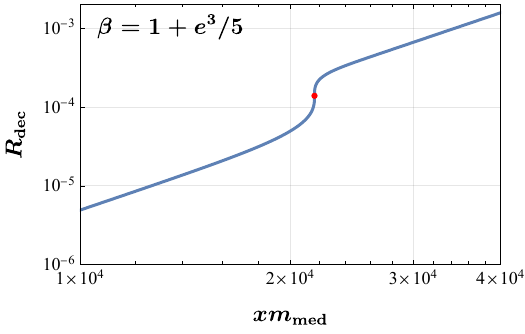}
    \includegraphics[width=0.325\linewidth]{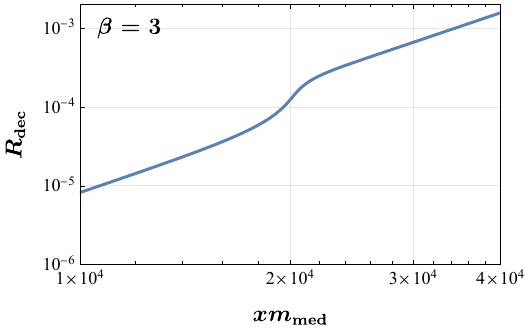}
    \caption{The scale factor at decoupling $\Rdec$ for a nearest neighbor MDM pair as a function of their initial comoving separation $x$. Here, $\mdm=10^{-4}M_\odot$, $\mmed=10^{-20}~\text{eV}$, and $\GpoG$ is varied as labeled in each panel. The blue solid lines show the analytic result from~\eqref{eq:decoupling} with $F$ given by~\eqref{eq:force_balance_real}, while the red dashed line in the left panel shows the physical decoupling scale factor when the analytic result is multi-valued. 
    The dotted red curve corresponds to the step-function approximation in Eq.~\eqref{eq:Rdec} with the multi-valuedness resolved as prescribed in the text. } 
    \label{fig:a0_x_numeric}
\end{figure}

In Eq.~\eqref{eq:a0_DF}, one can immediately observe a discontinuity in the value of $a_0$ as $x$ go across the boundary $x=l_{\rm th}$. This is due to the two decoupling ``chances'' discussed above.
It may appear that this gap is a mathematical relic of the approximation of an abrupt stepwise boundary of $x\Rdec=\mmed^{-1}$ in Eq.~\eqref{eq:Rdec}, but it is in fact physical for sufficiently large $\GpoG$. Comparing the continuous full expression of the force~\eqref{eq:force_balance_real} between the binary MDM with the Hubble dragging in Eq.~\eqref{eq:decoupling}, the mapping from $x$ to $a_0 \propto x\Rdec$ can be multi-valued for $\GpoG > \GpoGbarrier$, as shown in Fig.~\ref{fig:a0_x_numeric}. 
The one-to-one mapping given by the dashed red line in the left panel, which is well approximated by \eqref{eq:Rdec} as shown by the red dotted line, describes the physical values at which decoupling occurs. Because a binary with given $x$ will decouple at the earliest possible time, the small-$x$ regime decouples under the influence of the DF, while the large-$x$ regime decouples under gravity alone.
Even for $\GpoG\leqslant \GpoGbarrier$, one can see from the right panel of Fig.~\ref{fig:a0_x_numeric} that the value of $a_0$ changes swiftly around $x=l_{\rm th}$. This suggests that the corresponding $a_0$ values will have low probability to arise in the distribution function $P$ in Eq.~\eqref{eq:Pxy}, which effectively serves as a gap in $a_0$.
As a result, we use Eq.~\eqref{eq:a0_DF} even though slightly more accurate results are available by using~\eqref{eq:force_balance_real} in~\eqref{eq:decoupling}, as~\eqref{eq:a0_DF} not only describes the gap in $a_0$ but also provides better analytic control.

The semi-minor axis $b_0$ is estimated by $(\text{tidal acceleration})\times(\text{free-fall time})^2$~\cite{Nakamura:1997sm}. The tidal forces that determine the orbital eccentricity are dominated by the next-nearest-neighbor MDM with comoving distance from the binary center $y>x$ (although see \cite{Raidal:2017mfl,Raidal:2018bbj,Vaskonen:2019jpv,Hutsi:2020sol} for further possible refinements). Due to the finite range of the DF, the tidal acceleration is
\begin{align}
a_T=\begin{cases}
    \dfrac{\GpoG G (x \Rdec) \mdm}{(y \Rdec)^3}\,, & y \Rdec<\dfrac{1}{\mmed}\vspace{2mm}\\
    \dfrac{G (x\Rdec) \mdm}{(y \Rdec)^3}\,, & y \Rdec>\dfrac{1}{\mmed}
\end{cases}\,.
\end{align}
The squared free-fall time at large initial separation is dominated by the time spent outside the DF range and can thus be approximated by
\begin{align}
t^2_\text{ff}=\begin{cases}
    \dfrac{(x \Rdec)^3}{\GpoG G \mdm}\,, & (x\Rdec) < \dfrac{1}{\mmed}\vspace{2mm}\\
    \dfrac{(x \Rdec)^3}{G \mdm}\,, & (x\Rdec) > \dfrac{1}{\mmed}
\end{cases}\,.
\end{align}
The two expressions together with (\ref{eq:a0_DF}) imply that
\begin{align}\label{eq:b0_DF}
b_0=c_2\times \begin{cases}
    \dfrac{x^3}{y^3}a_0 \, , & y>x>l_{\rm th}\vspace{2mm} \, , \\
    \dfrac{1}{\GpoG}\dfrac{x^3}{y^3}a_0 \, , & y>l_{\rm th}>x\vspace{2mm} \, , \\
    \dfrac{x^3}{y^3}a_0 \, , & l_{\rm th}>y>x \, .
\end{cases}
\end{align}
Like $c_1$, $c_2$ may depend on $x$ and $y$ but for simplicity is taken as constant $c_2=0.8$ \cite{Ioka:1998nz}.

\begin{figure}[t]
    \centering
    \includegraphics[width=0.45\textwidth]{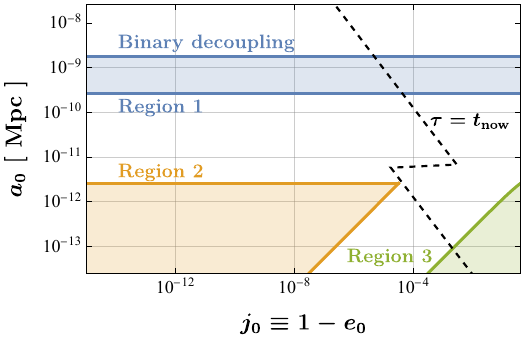} \hspace{3mm}
    \includegraphics[width=0.45\textwidth]{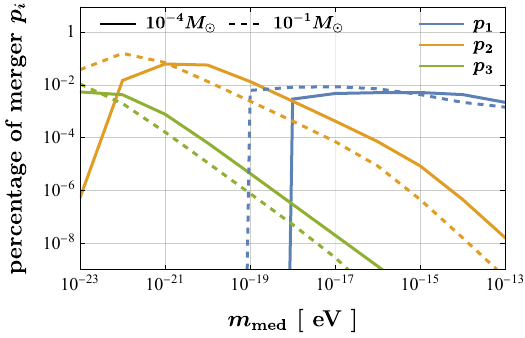}
    \caption{{\it Left:} An illustration of the three different regions of $(a_0,1-e_0)$, with $\mmed=10^{-18}{\rm eV} \approx (6 \times 10^{-12}~\text{Mpc})^{-1}$, $\GpoG=100$, and $\mdm = 10^{-4}M_\odot$. Binaries with a lifetime equal to the age of the universe are shown by the black dashed curve, with those that merge before today falling to the left of the curve. {\it Right:} Percentage $p_i$ of nearest-neighbor MDM pairs from Regions $i$ that decouple to form binaries before matter-radiation equality and merge by today, shown as a function of $\mmed$ with $\fdm=10^{-2}$. Solid (dashed) lines indicate $\mdm = 10^{-4}~(10^{-1})M_\odot$, and all other parameters are the same as in the left panel.}
    \label{fig:three_regions_cartoon}
\end{figure}

With the $a_0$ and $b_0$ in Eq.~\eqref{eq:a0_DF} and~\eqref{eq:b0_DF}, we may define a quantity
\begin{align}\label{eq:e0_definition}
e_0=\sqrt{1-\left(\dfrac{b_0}{a_0}\right)^2}\,.
\end{align}
In some regions of parameter space, this can be interpreted as the orbital eccentricity, but to see this we must first understand the parameter space of $(a_0,e_0)$.
The discontinuities of Eqs.~\eqref{eq:a0_DF} and~\eqref{eq:b0_DF} suggest that the distribution function $P$ will be divide into three disconnected regions when it is translated into $(a_0, e_0)$ from $(x, y)$ in~\eqref{eq:Pxy}.
These regions are illustrated in the left panel of Fig.~\ref{fig:three_regions_cartoon}, and can be interpreted as
\begin{itemize}
    \item Region 1: Binaries decouple under the influence of gravity only [top lines of (\ref{eq:a0_DF}) and (\ref{eq:b0_DF})].
    \item Region 2: Binaries decouple under the influence of the DF, but the tidal forces are only gravitational [bottom line of (\ref{eq:a0_DF}) and middle line of (\ref{eq:b0_DF})].
    \item Region 3: Binaries decouple under the influence of the DF and the next-nearest-neighbor MDM is also close enough for the DF to be effective in the tidal force [bottom lines of (\ref{eq:a0_DF}) and (\ref{eq:b0_DF})].
\end{itemize}
The lower bound for Region 1 and the upper bound for Regions 2 and 3 in $a_0$ are 
\begin{align}
\label{eq:a0_bound_region1}
a_0&> \dfrac{\GpoG c_1}{\mmed}\,, &(\text{Region 1})
\\
\label{eq:a0_bound_regions23}
a_0&< \dfrac{c_1}{\mmed}\,. &(\text{Regions 2 \& 3})
\end{align}
Additional boundaries for Regions 2 and 3 are given by
\begin{align}
\label{eq:decoupling_a_1}
a_0 &= \dfrac{c_1(1-e^2_0)^{2/3}\GpoG^{4/3}}{c^{4/3}_2\mmed},\quad & e_0> \sqrt{1-\dfrac{c^2_2}{\GpoG^2}} & \,,& (\text{Region 2})\vspace{2mm}\\
\label{eq:decoupling_a_23}
a_0 &= \dfrac{c_1(1-e^2_0)^{2/3}}{c^{4/3}_2\mmed},\quad & e_0< \sqrt{1-c^2_2}\,. & \quad & (\text{Region 3})
\end{align}
Also, there is an upper bound on $a_0$ because decoupling must occur before matter radiation equality, as discussed above (\ref{eq:Rdec}). Specifically, $x < \fdm^{1/3} \bar{x}$ for Region 1 and $x < (\fdm \GpoG)^{1/3} \bar{x}$ for Regions 2 and 3, which leads to
\begin{align}
\label{eq:a0_bound_decouple_region1}
a_0& < c_1\fdm^{1/3} \bar{x} \,, &(\text{Region 1})
\\
\label{eq:a0_bound_decouple_regions23}
a_0& < c_1\GpoG^{1/3} \fdm^{1/3} \bar{x} \,. &(\text{Regions 2 \& 3})
\end{align}
Comparing (\ref{eq:a0_bound_decouple_region1}) to (\ref{eq:a0_bound_region1}), binaries will only form in Region 1 if $(\mmed / (6 \times 10^{-23}~\text{eV})) (\mdm/M_\odot)^{1/3} > \beta$; otherwise, all binaries decouple under the influence of the DF.

Now let us return to the physical interpretation of $e_0$ in (\ref{eq:e0_definition}) and its relationship to the binary merger time. 
In Region 1, $e_0$ should be identified as $e_{\rm GR}$, even though the orbit may or may not be Keplerian depending on $\rmin$: when $\rmin>\mmed^{-1}$, $e_0$ describes the eccentricity of the whole orbit of the GR scenario of Sec.~\ref{sec:config1}; and when $\rmin<\mmed^{-1}$ as in Sec.~\ref{sec:config2}, $e_0$ describes only the portion of the orbit with $r\gtrsim\mmed^{-1}$ (see Appendix~\ref{sec:params_ae}).
In Regions 2 and 3, because of~\eqref{eq:a0_bound_regions23} and $c_1=0.4$, the DF is within range for the entire orbit. Therefore, $e_0$  should be identified as the orbital eccentricity in either the $\GpoG{\rm GR}$ scenario in Sec.~\ref{sec:config3} or the DF scenario in Sec.~\ref{sec:config4}. 
The corresponding binary lifetime can be estimated by matching to the four configurations in Sec.~\ref{subsec:evo_DF}. For the parameter choice in the left pane of Fig.~\ref{fig:three_regions_cartoon}, the binaries that merge before today are those that fall to the left of the black dashed curve. The segments of this curve are determined, from top to bottom, by $\taumix$ in (\ref{eq:tau_mixedrange}) and $\tau_{\GpoG \text{GR}}$ in (\ref{eq:tau_betaGR}), with the discontinuity located at $a_0 = \mmed^{-1}$. Note that the boundary $\mmed^{-1}$ therein always resides in the gap between Region 1 and Regions 2/3 regardless of parameter choice. 

Using these results, the fraction of binaries in the three regions can be calculated.
The right panel of Fig.~\ref{fig:three_regions_cartoon} shows the fraction $p_i$ of nearest-neighbor MDM pairs that can decouple before matter-radiation equality and merge before today, which can serve as a reference of the dominant contributing region to the SGWB. As expected, Region 1 (3) dominates at large (small) $\mmed$ where the DF has a very short (long) effective range, while Region 2 dominates at intermediate $\mmed$. Note that binaries do not form at all in Region 1 when either $\mmed$ or $\mdm$ is too small, as remarked upon below (\ref{eq:a0_bound_decouple_region1}). The peak probability of $p_2$ is larger than the peak of $p_3$ in part because binaries in Region 2 have larger eccentricity and thus shorter lifetime than those in Region 3, allowing more Region 2 binaries to merge before today. Because not all nearest-neighbor pairs decouple, and also not all of those that decouple have merged, $p_1+p_2+p_3$ is less than unity and varies with varying parameters.

Clearly, the parameter space is more complicated than that in the cases of a massless or infinitely massive DF mediator. The former has all binaries in Region 3 and all binary lifetimes determined by $\tau_\text{DF}$, while the latter has all binaries in Region 1 and all binary lifetimes determined by $\tau_\text{GR}$ (though the boundaries of these regions adjust with the mediator mass). The next section will show how these complications lead to a variety of features in the SGWB. A general statement that can be made is that, similar to the massless mediator case, binaries with a (not too) massive mediator can decouple earlier and have shorter merger lifetimes. This can enhance the binary merger rate compared to the gravity-only case. Additionally, binaries that decouple in Region 2 are much more eccentric than those that decouple in Region 3 because the tidal force is smaller in Region 2. This also has the effect of shorter merger lifetimes and a higher merger rate compared to the massless-mediator case. As we will see, an enhanced binary merger rate can correspondingly increase the SGWB up to a point. But for some parameter space, binaries can merge too early and instead suppress the SGWB.

\section{SGWB with a finite mediator mass}
\label{sec:SGWB}

With all the ingredients, the SGWB spectrum can be calculated as
\begin{align}\label{eq:SGWB}
\Omega_{\rm GW}(\fGW)=\dfrac{\fGW}{\rho_c} \int de_0\, d a_0 \dfrac{n_{\MDM}}{4} P(a_0,e_0)\dfrac{dE_{\rm GW}}{d\fGWs}\left[(1+z(t(a_0,e_0)))\fGW\right]\,,
\end{align}
where $f_{\rm GW}$ is the GW frequency observed today, $P$ is the binary distribution~\eqref{eq:Pxy} after changing the variables from $(x, y)$ to $(a_0, e_0)$ properly on its three disconnected regions, $t=t_{\rm dec}+\tau$ is the time when the binary merges dependent on its initial conditions $(a_0,e_0)$, and $t_{\rm dec}$ is the cosmic time when the binary decouples, which corresponds to $R_{\rm dec}$ discussed in Sec.~\ref{sec:distribution}.
The observed GW frequency $\fGW$ is related to the emitted GW frequency in the source frame $\fGWs$ as $\fGWs=(1+z)\fGW$, with $z(t)$ the redshift at the merger time $t$, so the highest frequency in the SGWB is exactly $\fmaxs$. For a given $f_{\rm GW}$, the boundaries of the integration are determined by the boundaries of the decoupling regions discussed in Eqs.~(\ref{eq:a0_bound_region1})--(\ref{eq:a0_bound_decouple_regions23}), the boundaries of $dE_\text{GW}/d\fGWs$ in (\ref{eq:dEdfGWs_analytic}) beyond which the emission is zero, and a constraint on the merger time. The merger time $t$ should satisfy $t_{\rm form}<t(a_0,e_0)<t_{\rm now}$, where $t_{\rm now}$ is the cosmic time today and $t_{\rm form}$ is the formation time of the MDM. Note that $t_{\rm form}$ is model dependent. For formation mechanisms based on a cosmic phase transition, we assume the cosmic temperature $T_{\rm form}$ at MDM's formation may be comparable to $\rho^{1/4}_\MDM$~\cite{Bai:2018dxf,Bai:2018vik,Bai:2022kxq}. We also assume radiation domination between $t_{\rm form}$ and $t_\text{eq}$.
The other side of this constraint, $t<t_{\rm now}$, includes only those binaries that have already merged in the integration. This is a good approximation because binaries spend most of their inspiral time very close to their initial conditions, and thus the vast majority of unmerged ones only emit at very low frequencies outside the range of interest.
For a similar reason, the redshift factor $z(a_0,e_0)$ in~\eqref{eq:SGWB} is evaluated at the time of the merger rather than the time the GWs are emitted, as evolution at higher (more detectable) frequencies proceeds relatively quickly in comparison.

As discussed in Sec.~\ref{sec:distribution}, in Region 1 of $P$, $a_0$ and $e_0$ describe the gravity-only part of the binary orbit, and hence are related to the initial total energy $E_0$ and individual angular momentum $J_0$ of the binary system via
\begin{align}
E_0=-\dfrac{G\mdm^2}{2a_0}\,,\quad J_0=\dfrac{\mdm}{2}\left(\dfrac{a_0(1-e^2_0)G\mdm}{2}\right)^{1/2}\,.
\end{align}
While in Region 2 and 3 the DF is active, and therefore $E_0$ and $J_0$ are calculated by
\begin{align}
E_0=-\dfrac{\GpoG G\mdm^2}{2a_0}\,,\quad J_0=\dfrac{\mdm}{2}\left(\dfrac{a_0(1-e^2_0)\GpoG G\mdm}{2}\right)^{1/2}\,.
\end{align}
The merger time $t$ and the corresponding $dE_{\rm GW}/d\fGWs$ can then be estimated with the emission spectrum and merger lifetime estimated in Secs.~\ref{subsec:evo_DF} and \ref{subsec:GW_emi_spec_approx}.

\subsection{Features in the SGWB spectrum}

\begin{figure}[t!]
    \centering
    \includegraphics[width=0.95\linewidth]{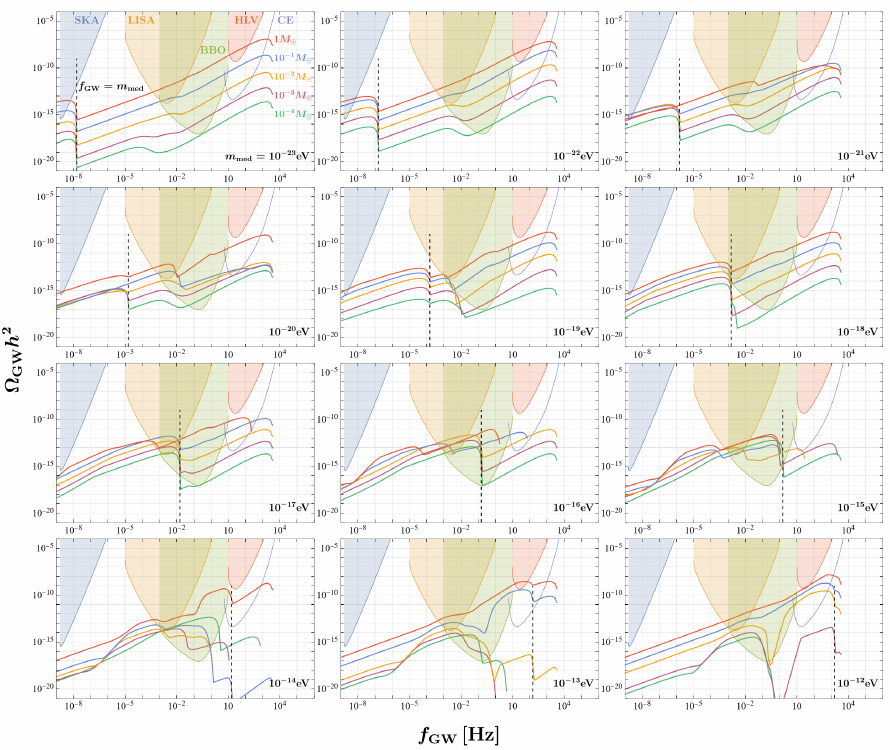}
    \caption{
    The calculated SGWB spectra for various $\mdm$ and $\mmed$, with $\rhodm=(0.1~\GeV)^4$, $\GpoG=100$, and $\fdm=10^{-2}$. The different panels corresponds to $\mmed=10^{-23}$--$10^{-12}$~eV from top-left to bottom-right, and in each panel the SGWB for $\mdm=10^{-4}$--$1\uu M_\odot$ are presented.
    Also shown are the power-law integrated sensitivity of the Square Kilometer Array (SKA)~\cite{Carilli:2004nx}, LISA~\cite{LISA:2017pwj}, the Big Bang Observer (BBO)~\cite{Corbin:2005ny}, the (Hanford and Livingston) LIGO-Virgo network (HLV)~\cite{LIGOScientific:2014pky,VIRGO:2014yos} and the Cosmic Explorer (CE)~\cite{Reitze:2019iox}, taken from~\cite{Schmitz:2020syl}.
    }
    \label{fig:Omegah2_vs_mmed_f2}
\end{figure}

\begin{figure}[t]
    \centering
    \includegraphics[width=0.95\linewidth]{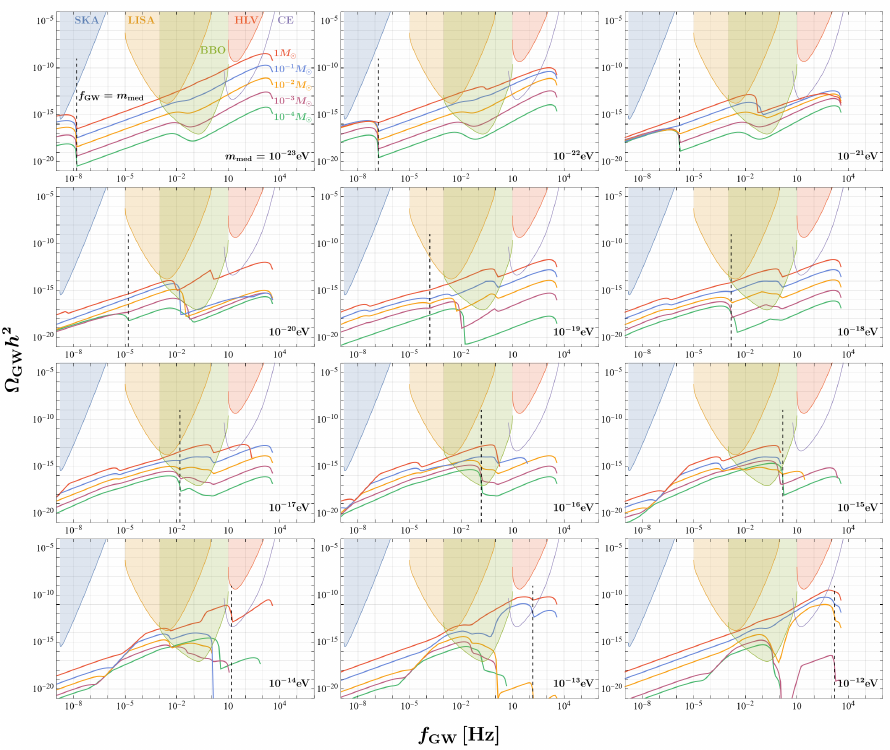}
    \caption{Same as Fig.~\ref{fig:Omegah2_vs_mmed_f2}, but with $\fdm=10^{-3}$.}
    \label{fig:Omegah2_vs_mmed_f3}
\end{figure}

In Fig.~\ref{fig:Omegah2_vs_mmed_f2} we show the calculated SGWB for $\mmed=10^{-23}$--$10^{-12}~{\rm eV}$, $\fdm = 10^{-2}$, and various different $\mdm$, compared with the reach of various present and future GW observatories. Fig.~\ref{fig:Omegah2_vs_mmed_f3} is identical, except with $\fdm=10^{-3}$.
Due to the complications in the binaries' initial distributions, lifetimes, and emitted GW spectra stemming from the finite range of the DF, the SGWB becomes more ``ragged'' compared with the gravity-only and massless-DF cases~\cite{Bai:2023lyf}.
There are some common features shared by most spectra across the range of $\mmed$ and $\mdm$ examined: the SGWB has a plunging knee at around $\fGW=\mmed$ (shown by the dashed vertical lines in each panel) and a softer one at $\fGW=f_\GpoG$ (which depends on $\mdm$). These are clearly inherited from the features of $dE_\text{GW}/d\fGWs$, where in particular the relative enhancement in the middle piece of Eq.~\eqref{eq:dEdfGWs_analytic} is important for making some of the example spectra detectable.
Some features, on the other hand, are presented only in a specific range of $\mmed$. 

For concreteness, for the remainder of this subsection let us exclusively examine $\mdm=0.1M_\odot$ and $\fdm = 10^{-2}$, the blue curves (the second from the top in each panel) in Fig.~\ref{fig:Omegah2_vs_mmed_f2}. 
As $\mmed$ increases from $10^{-22}$~eV to $10^{-20}$~eV, the amplitude of the SGWB at $\fGW\gtrsim 0.1$~Hz decreases, the knee at $\fGW\sim\mmed$ gradually disappears, and a new one develops at $\fGW\sim 10^{-2}$~Hz.
When $\mmed$ is heavier than $\sim 10^{-16}$~eV, on the other hand, the SGWB either cuts off at a lower frequency or shows a clear deficit at around the maximum frequency.
Analogous behaviors also show up on curves of other $\mdm$ and $\fdm$, though sometimes at different $\mmed$ and $\fGW$.
These complications result from the discontinuities of the binaries' initial distribution and lifetime, which we explain below.

The features in the mass range $10^{-22}~{\rm eV}\lesssim \mmed\lesssim 10^{-20}~{\rm eV}$ are mainly explained by an examination of Region 2, because the majority of binaries that can merge by today come from this region, as seen in Fig.~\ref{fig:three_regions_cartoon}. There, the spectral amplitude at frequencies larger than 0.1~Hz decreases at larger mediator mass because the mergers happen too early on and cannot contribute to the highest frequencies. 
To elaborate, the upper bound in $a_0$ and the lower bound in $e_0$ suggest a maximum lifetime for binaries residing in this region.
For instance, if the merger lifetime is determined by $\tau_{\GpoG{\rm GR}}$ as in Configuration 3 of Sec.~\ref{sec:config3} (suggested by the upper bound $a_0\leqslant c_1/\mmed$ of this region), the merger lifetime is bounded as
\begin{equation} \label{eq:tau_max_r2}
    \tau\lesssim (3c^4_1c^7_2)/(170\GpoG^9 G^3\mmed^4\mdm^3) \, , \quad \quad \quad \quad \text{(Region 2),}
\end{equation}
as estimated by the leading order behavior of Eq.~\eqref{eq:tau_betaGR}. 
Therefore, when $\mmed$ increases, the number of late-merging binaries in Region 2 decreases, and the contributions to the SGWB are more redshifted to the lower-frequency region, explaining the decrease in the spectral amplitude at $\fGW\gtrsim 0.1$Hz as $\mmed$ increases in this range. When $\mmed$ is too heavy, the knee at $\fGW\sim\mmed$ disappears from being redshifted away because all binaries formed in Region 2 merge before today. Specifically, for some binaries to have a merger time $t\approx\tau>t_{\rm now}$ requires
\begin{align} \label{eq:region2_no_merge_today}
\mmed<4 \times 10^{-21}{\rm eV}\left(\dfrac{\mdm}{0.1M_\odot}\right)^{-\frac{3}{4}}\left(\dfrac{\GpoG}{100}\right)^{-\frac{9}{4}}\,.
\end{align}
For a mediator mass above this cutoff but below the value where region 1 binaries can be formed, the SGWB at high frequencies is contributed solely by region 3 binaries.

The knee at $\fGW\sim 10^{-2}$~Hz when $\mmed=10^{-21}$--$10^{-19}$eV is also related to the lifetime upper bound discussed above.
Note that in Eq.~\eqref{eq:SGWB} the redshift $z$ of the spectrum is determined not by the binary lifetime $\tau$ alone, but by the cosmic merger time $t=\tau+t_{\rm dec}$, and it is possible to have $t\sim t_{\rm dec}\gg \tau$.\footnote{This possibility doesn't invalidate our discussions in the previous paragraph. To contribute to the high frequency part of $\Omega_{\rm GW}$, it must be that $t\approx\tau\gg t_{\rm dec}$ so that the GWs are not redshifted away.}
As the maximum possible lifetime $\tau$ in (\ref{eq:tau_max_r2}) decreases with respect to the increase of $\mmed$, a larger fraction of the binaries in Region 2 will have a cosmic merger time $t\approx t_{\rm dec}$, which is independent of $e_0$ and has a less strong dependence on $a_0$ than $\tau$.
This implies a ``pile-up'' in the spectrum.
Specifically, for $\mmed=10^{-20}$eV it can be found that $t_{\rm dec}$ dominates $\tau$ in merger time $t$ when $\tau\lesssim 10^7$s, corresponding to a redshift of $z\sim 10^6$, exactly the ratio between the spectral cut-off $f_{{\rm max},s}\sim 10^4~$Hz and the knee at $\fGW\sim 10^{-2}$~Hz.

\begin{figure}[t]
    \centering
    \includegraphics[width=0.32\linewidth]{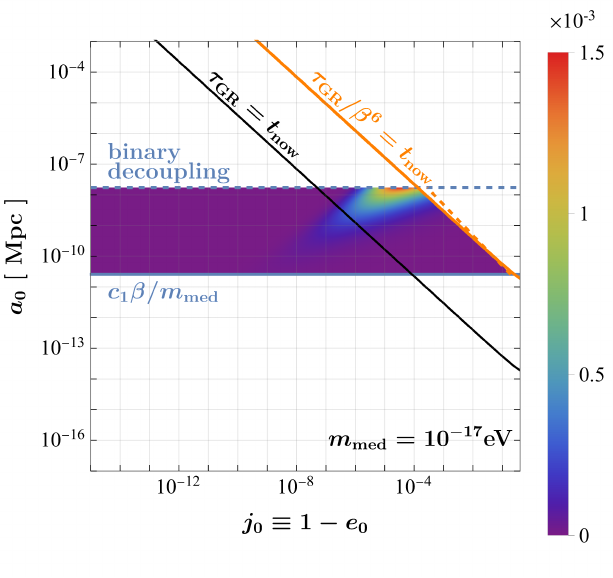}
    \includegraphics[width=0.32\linewidth]{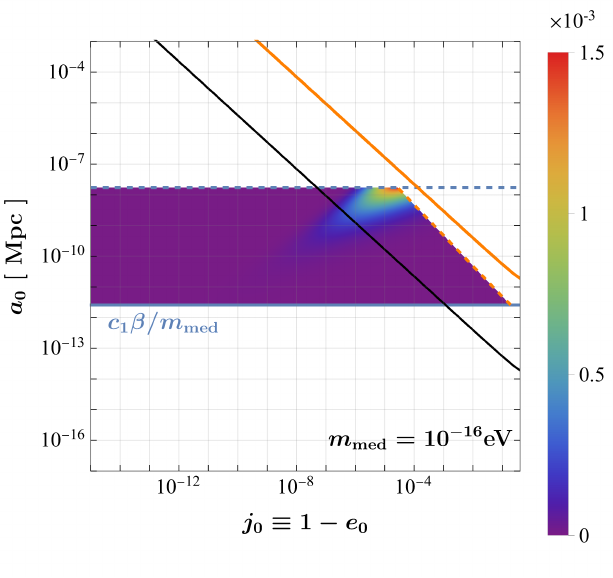}
    \includegraphics[width=0.32\linewidth]{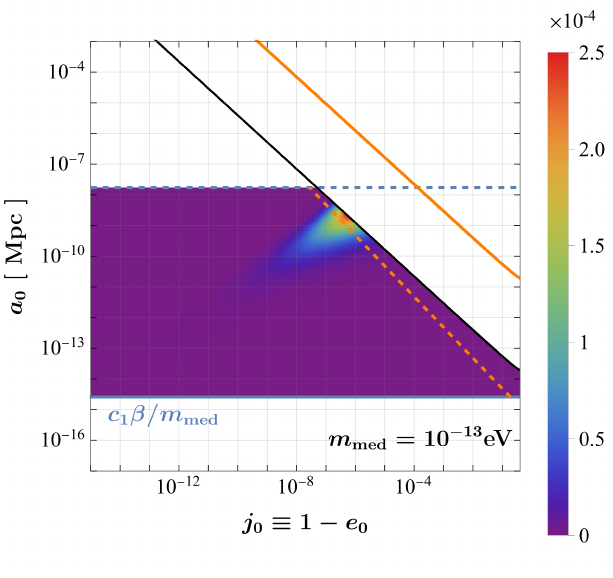}
    \caption{The heat maps of the probability distribution function $a_0(1-e_0)\uu P(a_0,e_0)$ in Region 1 for $\GpoG=100$, $\mdm=10^{-1}M_\odot$ and $\fdm=10^{-2}$, where the factor $a_0(1-e_0)$ is multiplied to compensate for the logarithmic plotting scheme. From left to right, the three panels have $\mmed=10^{-17}$, $10^{-16}$ and $10^{-13}$eV, respectively. The decoupling constraint is given by the dashed blue line, and the merger lifetime constraint is accounted for by the two orange lines and the black line (see the main text for more detailed explanation). }
    \label{fig:heatmap}
\end{figure}

We now turn to the spectral irregularity at $\mmed\gtrsim 10^{-16}$eV, which is more related to the influence of $U_{\rm eff}$ on the binaries' lifetimes.
For this range of $\mmed$, binaries in Region 1 of the initial distribution contribute dominantly to the SGWB, as shown in the right panel of Fig.~\ref{fig:three_regions_cartoon}.
And according to the discussion in Sec.~\ref{subsec:evo_DF}, the binaries' initial conditions may correspond to Configuration 1 or 2, with their lifetimes approximated as $\tau\approx\tau_{\rm GR}$ or $\tau\approx\taumix$, respectively.
With the binaries' decoupling constraint and their lifetime constraint taken into account, in Fig.~\ref{fig:heatmap} we show probability heat maps of valid parameter space in Region 1 that can contribute to the SGWB for $\mmed=10^{-17}$, $10^{-16}$, and $10^{-13}$eV.
The lifetime constraints of the two categories are shown as the solid black and orange curves, to the left of which the binaries can merge by today.
However, the binaries with $\tau\approx\taumix$ can only exist where either $U_{\rm eff}$ has no barrier, or the energy of the binary is large enough such that it can traverse the barrier, see the discussions in Sec.~\ref{sec:evo}.
The corresponding bounds on the initial $E$ and $J$ of the binaries, after translating to $a_0$ and $j_0=1-e_0$ with Eq.~\eqref{eq:EJtoae}, are shown as the dashed orange line, only to the left of which can the binaries' lifetime be estimated as $\tau\approx\taumix$.
It is then immediately clear why the SGWB with $\mmed= 10^{-16}$~eV loses its high frequency region: with the dashed orange curve standing to the left of the solid one, all the valid binaries in this occasion have their lifetime $\tau\ll t_{\rm now}$ and hence have their GWs redshifted to lower $\fGW$. The binaries between the solid and dashed orange lines instead become trapped in the gravity-only side of the barrier (as in the orange line of Fig.~\ref{fig:dEGWdf_numeric}) and have lifetimes determined by $\tau_\text{GR}$, which by comparison to the black line is much longer than the age of the universe so that they do not contribute appreciably to the SGWB today. By contrast, for $\mmed\sim 10^{-17}$~eV, the dashed orange line is to the right of the solid orange line throughout region 1, so binaries can merge as late as today and the high-frequency cutoff in the SGWB does not experience significant redshifting.
For $\mmed= 10^{-13}$~eV, on the other hand, binaries between the dashed orange and the solid black line have their lifetime determined as $\tau\approx\tau_{\rm GR}$.
As a result, these binaries can still have $\tau\approx t_{\rm now}$, and the SGWB extends all the way to $\fmaxs$.
The presence of multiple knees and ankles in the SGWB below the maximum $\fGW$ is the result of the superposition of the contributions from binaries merging with lifetimes of $\taumix$ and $\tau_{\rm GR}$.

\begin{figure}[t]
    \centering
    \includegraphics[width=0.5\linewidth]{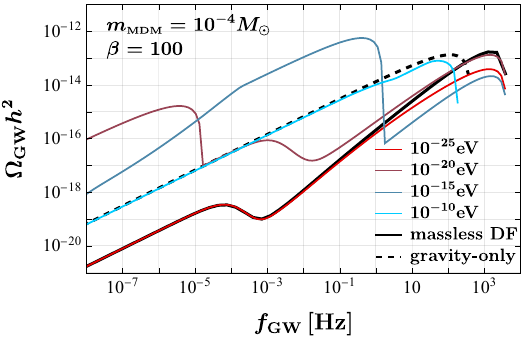}
    \caption{A collection of the SGWB for $\GpoG=100$, $\mdm=10^{-4}M_\odot$, and $\fdm=10^{-2}$. The SGWB from gravity-only (dashed black) and massless DF mediator (solid black) scenarios are shown as benchmarks, to be compared with the results for $\mmed=10^{-25}$--$10^{-10}$eV (colored curves).
    It is clear that as in the heavy and light $\mmed$ limit, the massive DF mediator results restore the gravity-only and massless DF ones.\protect\footnotemark
    }
    \label{fig:Omegah2_vs_mmed_2}
\end{figure}
\footnotetext{The deficit around the cut-off frequency for the $\mmed=10^{-10}$eV curve compared with the gravity-only case is caused by the gap between $f_{\rm trans}$ and $f_{\GpoG}$ in Eq.~\eqref{eq:dEdfGWs_analytic}, instead of the $\tau_{\rm GR}$ vs. $\tau_{\rm mix}$ difference discussed in Fig.~\ref{fig:heatmap}. For a slightly heavier $\mmed$ the calculated SGWB overlaps with the gravity-only result completely.}

Despite the drastic change in the spectra at intermediate $\mmed$, the SGWB still converges to the benchmark scenarios of gravity-only and massless-DF in the heavy- and light-mediator limits, as is shown in Fig.~\ref{fig:Omegah2_vs_mmed_2} where $\beta = 100$, $\fdm=10^{-2}$, and $\mdm=10^{-4}M_\odot$ are assumed.

\subsection{Spectral shapes and amplitudes}

\begin{figure}[t]
    \centering
    \includegraphics[width=0.55\linewidth]{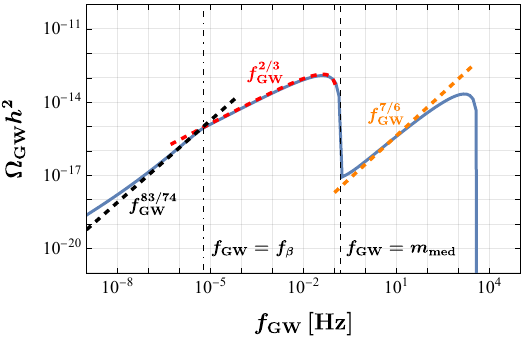}
    \caption{The SGWB for $\GpoG=100$, $\mdm=10^{-4}M_\odot$, $\mmed=10^{-16}$~eV, and $\fdm=10^{-2}$ (blue curve), compared with the spectral shape at $f_{\GpoG}<\fGW<\mmed$ and $\fGW\lesssim f_{\GpoG}$ analyzed as in the main text (dashed red and dashed black, respectively). The power-law behavior inferred from the massless DF scenario at large $\fGW$ is also shown as the dashed orange curve.}
    \label{fig:spectral_explanation}
\end{figure}

Let us turn to analytic approximations for the integral in (\ref{eq:SGWB}) to provide insight into and validate the SGWB spectra we obtained. To perform these approximations, it is easiest to focus on a choice of parameters and GW frequencies where the dominant contribution to the SGWB comes from a particular region of the initial distribution $P$ and the emission spectrum $dE_{\rm GW}/d\fGWs$. Take the example of $\mdm=10^{-4}M_\odot$ and $\mmed=10^{-16}{\rm eV}$ plotted in Fig.~\ref{fig:spectral_explanation}, where the binaries mainly come from Region 1 of $P$. 
Contributions to the SGWB from slightly below $\fGW=f_{\GpoG}$ up to $\fGW=\mmed$, including two knees at $\fGW=f_{\GpoG}$ and $\mmed$, predominantly come from the frequency range $f_{\GpoG}<\fGWs<\mmed$ of the source emission spectrum $dE_{\rm GW}/d\fGWs$ in Eq.~\eqref{eq:dEdfGWs_analytic}.
For these features to appear, the mergers must be relatively recent on cosmological scales so as not to redshift the features away. Therefore, the relevant emission spectrum and merger lifetime are well approximated by
\begin{align}\label{eq:dEdfGWs_middle}
\dfrac{dE_{\rm GW}}{d\fGWs}&=\left(\dfrac{\pi^2 \GpoG^2 G^2\mdm^5}{54\fGWs}\right)^{1/3}\Theta\left((1+z)f_{\rm GW}-f_\GpoG\right)\Theta\left(\mmed-(1+z)f_{\rm GW}\right)\,,\\
1+z&=\left(\dfrac{2}{3H_0\sqrt{\Omega_M}t}\right)^{2/3}\,,\quad t\approx\tau\approx\dfrac{3a^4_0(1-e^2_0)^{7/2}}{170\GpoG^6 G^3 \mdm^3}\,.
\end{align}
With these, the integration in Eq.~\eqref{eq:SGWB} can be performed analytically, which gives
\begin{align}\label{eq:Omega_shape_full}
\Omega_{\rm GW}&=A\left[\Phi(B_{\rm up})-\Phi(B_{\rm down})\right]\,,\\
\label{eq:func_Phi}
\Phi(B)&=\dfrac{27}{101}\left(B^{\frac{1616}{999}}\Gamma\left(\frac{58}{37},\frac{B}{c^{37/16}_2}\right)-c^{101/27}_2\Gamma\left(\frac{86}{27},\frac{B}{c^{37/16}_2}\right)\right)\,,\\
A&=\dfrac{\pi^{2/3}(\fdm\uu H^2_0 c^8_1\bar{x}^8\Omega_M)^{1/9}}{\GpoG^{2/3}2^{25/9}3^{5/9}85^{2/9}c^{59/27}_2}\Omega_{\rm DM}f^{2/3}_{\rm GW}\,.\label{eq:factor_A}
\end{align}
where $\Gamma(a,z)$ is the upper incomplete gamma function. $B_{\rm up}$ and $B_{\rm down}$ are related to the upper and lower bound of the $a_0$ integration.
For $f_{\GpoG}<f_{\rm GW}<\mmed$, the upper bound of $a_0$ is not determined by the Heaviside functions $\Theta$ in~\eqref{eq:dEdfGWs_middle}, but by the requirement of $t\approx\tau_{\rm GR}<t_{\rm now}$.
The corresponding $B_{\rm up}$ and $B_{\rm down}$ are then 
\begin{align}\label{eq:B1}
B_{\rm up}&=c_2\left(\dfrac{170 \fdm^4 \GpoG^6 G^3\mdm^3 t_{\rm now}}{3 c^4_1 \bar{x}^4}\right)^{\frac{3}{16}}\,,\\
B_{\rm down}&=c_2\left(\dfrac{340\fdm^4 \GpoG^6 G^3\mdm^3}{9c^4_1 H_0 \bar{x}^4\sqrt{\Omega_M}}\right)^{\frac{3}{16}}\left(\dfrac{f_{\rm GW}}{\mmed}\right)^{\frac{9}{32}}\,.
\end{align}
In this region the function $\Phi$ does not rely strongly on the $f_{\rm GW}$ factor in $B_{\rm down}$, and the SGWB thus has a power-law behavior of $f^{2/3}_{\rm GW}$, coming from the factor $A$ in Eq.~\eqref{eq:factor_A}.

For the spectrum at $f_{\rm GW}<f_\GpoG$, the upper bound of the $a_0$ integration is set by the Heaviside function in~\eqref{eq:dEdfGWs_middle}, and the corresponding $B_{\rm up}$ is
\begin{align}\label{eq:B2}
B_{\rm up}=c_2\left(\dfrac{340\fdm^4\GpoG^6 G^3\mdm^3}{9c^4_1 H_0 \bar{x}^4}\right)^{\frac{3}{16}}\left(\dfrac{f_{\rm GW}}{\mmed}\right)^{\frac{9}{32}}\left(\dfrac{\pi^2}{2\GpoG G \mdm\mmed}\right)^{\frac{9}{64}}\,.
\end{align}
For $f_{\rm GW}<f_\GpoG$, the value of $B$ is small enough such that we may take $\Gamma(a,z)\approx \Gamma(a)$ in~\eqref{eq:func_Phi}, which simplifies the spectrum to
\begin{align}\label{eq:OmegaGW_piece_low}
\Omega_{\rm GW}\approx \dfrac{3^\frac{68}{37}85^\frac{3}{37}\pi^{\frac{83}{74}}\Gamma\left(\frac{58}{37}\right)\fdm^\frac{49}{37}\GpoG^{\frac{137}{148}}\Omega_{\rm DM}}{404\times 2^{\frac{59}{148}}(c^{12}_1 c^{21}_2 H^3_0 \bar{x}^{12})^{\frac{1}{37}}\Omega^{\frac{3}{74}}_M}\left(\dfrac{G\uu\mdm}{\mmed}\right)^{\frac{101}{148}}\fGW^{83/74}\,.
\end{align}
The comparison between the SGWB calculated with~\eqref{eq:B1} and~\eqref{eq:B2} and the numeric results are shown in Fig.~\ref{fig:spectral_explanation}.

For $\fGW$ larger or smaller than the range discussed above, the main contributor to the SGWB is not as clear.
For example, the $dE_{\rm GW}/d\fGWs$ that can contribute to $\fGW>\mmed$, Eq.~\eqref{eq:dEdfGWs_DF}, has a varying power-law index that depends on the initial condition $a_0$ and $e_0$~\cite{Bai:2023lyf}.
For a massive DF mediator, 
$a_0$ and $e_0$ correspond to those describing the initial orbital configuration of a binary's Configuration 4 evolution (Sec.~\ref{sec:config4}), which usually have to be obtained numerically after evolving the binary through Configuration 1-3 and thus are not easily tracked.
The generated SGWB therefore is slightly different from the $\fGW^{7/6}$ power-law behavior in the massless DF mediator's case \cite{Bai:2023lyf}, as compared in Fig.~\ref{fig:spectral_explanation}.

At $\fGW\ll f_{\GpoG}$, the SGWB consists of both 
the $f_{\GpoG}<\fGWs<\mmed$ branch of those binaries that merged during radiation domination, as well as the $\fGWs<f_{\rm trans}$ branch of the ones that merged during matter domination. This latter branch was neglected in the analytic approximations above, resulting in a deficit in the analytic approximation at the smallest frequencies.
As $\fGW$ decreases, the spectral shape in the full numerical calculation starts to deviate from the $\fGW^{83/74}$ power law in~\eqref{eq:OmegaGW_piece_low} and gradually transitions to a $\fGW^{2/3}$ power law as in the gravity-only case.

\subsection{Visibility at GW experiments}
\label{subsec:exp_visibility}

The features in the SGWB brought by the finite range of the DF, specifically the knees in the spectra that are more or less related to the mediator mass, provide interesting targets for experiments to search for.
As shown in Fig.~\ref{fig:Omegah2_vs_mmed_f2}, if the MDM can make up one percent of all the DM, SKA may have a chance to distinguish signatures for $\mmed=10^{-23}$eV, and LISA may resolve $\mmed=10^{-16}$ or $10^{-17}$eV.
However, weak gravitational lensing observations \cite{Niikura:2017zjd,Smyth:2019whb,Griest:2013aaa,Macho:2000nvd,EROS-2:2006ryy,Wyrzykowski_2011,Niikura:2019kqi,Zumalacarregui:2017qqd,Oguri:2017ock,Mroz:2024wia,Croon:2020wpr,Bai:2020jfm} suggest a constraint of $\fdm\lesssim (\text{few}) \times 10^{-3}$ for (sub-)solar mass MDM candidates.
Therefore, Fig.~\ref{fig:Omegah2_vs_mmed_f3} suggests that the SGWB from MDM binaries may generally require GW observatories in the generation after those that are currently being utilized or constructed to be observed. Only a few choices of parameters lead to detectable signals at SKA or LISA, the observatories currently under construction. 

Additionally, there is a possibility for MDM with smaller masses, larger internal density, and/or larger $\GpoG$ than what is plotted in Figs.~\ref{fig:Omegah2_vs_mmed_f2} and \ref{fig:Omegah2_vs_mmed_f3} to be detectable by proposed high frequency experiments \cite{Li:2003tv,Aggarwal:2020olq,Ringwald:2020ist,Franciolini:2022htd,Domcke:2024mfu} while evading cosmic microwave background bounds on extra radiation degrees of freedom in the universe (both current \cite{Planck:2018vyg} and future \cite{Abazajian:2019eic}). This is because the the SGWB is more abundant today than at the time of recombination, since more binaries have merged between then and now. Such a possibility was demonstrated in \cite{Bai:2023lyf}, although the underlying theory for such large $\rhodm$ and $\GpoG$ is less motivated.

\section{Discussion and conclusions}
\label{sec:conclusion}

In this paper, we examined the formation, orbital evolution, and lifetime of MDM binaries, and the SGWB generated from their mergers in the presence of a finite-ranged attractive DF.
The range of the attractive DF is incorporated through the mass of its mediator, adding an exponentially decaying Yukawa force in additional to the inverse-squared gravitational force between the MDMs in the binary.
The Yukawa force could generate a barrier in the effective potential of a binary system and greatly complicates its orbital evolution and emitted GW spectrum.
Depending on the initial condition, at different stages the MDM binary may evolve as if it is only under the influence of gravity, as if the DF is active in strength but not radiating the DF mediator, as if the DF is long-ranged and its mediator can be emitted as radiation, or as a hybrid of these possibilities.
Correspondingly, the emitted GW spectrum from an individual binary is approximately piecewise, behaving in the same way as that of the gravity-only case at low GW frequency, acquiring an enhancement from the DF in the intermediate frequency regime when the DF mediator is not emitted, and acting like the spectrum with a massless DF mediator at high GW frequency after the DF mediator emission initiates.
The largest possible GW frequency emitted by the binary could be larger than that of the gravity-only case, depending on the mediator mass and the radius of the MDM.
The enhancement of the GW amplitude at intermediate frequencies and the increase in the maximum emitted frequency can boost the SGWB signal.

The initial distribution of the binary orbital configurations is also changed drastically by the finite-ranged DF due to the modification of the binaries' decoupling conditions and the influence from the next-to-nearest neighboring MDM.
Specifically, the distribution is generically divided into three disconnected regions. 
For different DF mediator masses, the binaries that merge by today come from different regions, and the binary merger rate could be enhanced compared to the gravity-only case as long as the DF mediator is not too heavy. 
But for some parameter space, binaries can merge too early and instead suppress the visibility of the emitted GW.

For MDMs that satisfy the abundance constraints from gravitational lensing, the SGWB resulting from MDM mergers could serve as a target for future GW detectors like SKA, LISA, BBO, CE, and high frequency detectors.
If a SGWB is detected, there are several possible ways to distinguish whether it results from dark binary mergers. First, the spectral shape itself differs from other SGWB sources like black hole mergers, phase transitions, and cosmic strings (see for example \cite{NANOGrav:2023hvm,Ellis:2023oxs,Sasaki:2018dmp,Caprini:2019egz,Hindmarsh:1994re}). This is particularly true for all of the sharp changes that are characteristic of mergers with a finite-mass DF mediator (for example, at $\fGW= f_\GpoG$ and $\mmed$). Second, it is likely that anisotropies in the SGWB \cite{NANOGrav:2023tcn} could distinguish between different progenitor models, though this is beyond the scope of this work. For example, the galactic component of the GW background could be disentangled from the full SGWB \cite{vanDie:2024htf}, and the distribution of MDM in the galaxy would be halo-like as opposed to astrophysical black holes concentrated in the galactic disk.  Finally, in addition to the SGWB, it is possible to detect the GWs from the inspiral or merger of individual dark binaries. With the presence of the DF, the GW waveforms would differ sharply compared to the case of ordinary black hole or neutron star mergers. This could be an interesting topic for future work in source modeling using numerical relativity techniques.

The radiated dark force mediators could contribute to the extra radiation degrees of freedom of the universe, which has been considered in \cite{Bai:2023lyf} and shown to be negligible for the parameter examples in this paper. On the other hand, some radiated mediators could become nonrelativistic and have a larger contribution to the total energy of the universe than $\Omega_{\rm GW}$ but still have a tiny contribution to the dark matter energy density.

Throughout this work, the emission processes from the MDM binary are treated in a period-averaged manner, with the threshold for the start of the DF emission set to be when the orbiting frequency is similar to the DF mediator mass. 
However, it may be possible for binaries with smaller orbital frequencies to radiate the DF mediator during a portion of their full orbits where the instantaneous angular frequency is large enough.
Such binaries should have orbits with $\rmin<\aemit <\rmax$, and the radiation should happen where $r_1<\aemit$.
On the other hand, a full numeric solution for the whole binary trajectory would be required to account for this possibility, which is much more computationally costly compared with the period-averaged treatment.

Also due to the computation cost, we simply take $\fGWs=2/T$ and do not include the higher harmonics of the GW emission. For a binary black hole merger under only gravity, it has been argued that the higher harmonics could be important when the binary orbit is eccentric~\cite{Maggiore:08textbook,Enoki:2006kj,vanDie:2024htf}.
As the binary orbit is not closed with the finite-ranged DF, the higher harmonics may be more important than those in the gravity-only and massless DF cases \cite{Bai:2023lyf}, which is left for future work.

\subsubsection*{Acknowledgments}
The work of YB is supported by the U.S. Department of Energy under the contract DE-SC-0017647. The work of SL is supported by the Area of Excellence (AoE) under the Grant No. AoE/P-404/18-3 issued by the Research Grants Council of Hong Kong S.A.R. The work of NO is partially supported by the National Science Centre, Poland, under research grant no.~2020/38/E/ST2/00243. We thank the Center for High Throughput Computing at the University of Wisconsin-Madison for providing computing resources~\cite{https://doi.org/10.21231/gnt1-hw21}.

\appendix

\section{Orbital parameter relationships} 
\label{sec:params_ae}

Let us elaborate on the quantities $a_{\rm GR}$ and $e_{\rm GR}$ in $\taumix$ of Eq.~(\ref{eq:tau_mixedrange}) from Configuration~2. For the other three configurations, the entire orbit is Keplerian, so $a$ and $e$ are well defined. However, Configuration~2 is when the orbit feels only the force of gravity near apogee (furthest separation) but feels the DF near perigee (closest separation). For this configuration, the orbit is non-Keplerian and does not follow a closed elliptical path, but has an apsidal precession similar to the well-known example of Mercury's orbit. 
The maximum and minimum orbital radii can be related to the orbital energy and angular momentum by 
\begin{align}
&r_{\rm max,mix}\approx -\dfrac{G\mdm^2}{2E}\,,\quad r_{\rm min,mix}\approx\dfrac{2J^2}{\GpoG G\mdm^3}\,,
\end{align}
where for convenience we define $r=2r_1$. By comparison, the gravity-only Configuration~1 or DF Configurations~3 and~4 have radii at apogee and perigee given by
\begin{alignat}{2}
&r_{\rm max,GR}\approx -\dfrac{G\mdm^2}{2E}\,, &&r_{\rm min,GR}\approx\dfrac{2J^2}{G\mdm^3}\,; \\
&r_{\rm max,DF/{\GpoG GR}}\approx -\dfrac{\GpoG G\mdm^2}{2E}\,,\quad &&r_{\rm min,DF/{\GpoG GR}}\approx\dfrac{2J^2}{\GpoG G\mdm^3}\,.
\end{alignat}
Notice that $r_{\rm max,mix} \approx r_{\rm max,GR}$ and $r_{\rm min,mix} \approx r_{\rm min,DF/{\GpoG GR}}$. These three sets of quantities can be used to define an effective ``semimajor axis'' $a \equiv (\rmax + \rmin)/2$ and ``ellipticity'' $e \equiv (r_{\rm max}-r_{\rm min})/(r_{\rm max}+r_{\rm min})\approx 1-2r_{\rm min}/r_{\rm max}$. For Configuration~2, the various definitions are related by (defining $j \equiv 1-e$)
\begin{align} \label{eq:eccentricity_mixedrange}
j_{\rm GR}\approx\GpoG j_{\rm mix}\approx \GpoG^2 j_{\rm DF/\GpoG GR}\,.
\end{align}
For the purposes of the Configuration 2 lifetime in Eq.~(\ref{eq:tau_mixedrange}), $e_\text{GR}$ should be used, and this is what results from $e_0$ in Eq.~(\ref{eq:e0_definition}). The correct $a_0$ is given in Eq.~(\ref{eq:a0_DF}) because when $j \ll 1$, $\rmax \gg \rmin$ so that $a \approx \rmax/2$ is well approximated with $\rmax = r_{\rm max,mix} = r_{\rm max,GR}$ alone.

\newpage
\section{Notation table}
\label{sec:notation_table}


    \centering
    \begin{tabular}{c|m{10.5cm}|c}
    \renewcommand{\arraystretch}{1.05}
         Symbol & \multicolumn{1}{c|}{Meaning} & Text placement  \\ \hline \hline
         $\mmed$ & Mass of the DF mediator. & Below Eq.~(\ref{eq:force_balance_real})\\
         $\GpoG$ & DF enhancement factor (as a multiple of $G$). & Below Eq.~\eqref{eq:force_balance_real}\\
         $\rhodm$ & Internal MDM density $=3 \mdm / (4 \pi \rdm^3)$. & Eq.~(\ref{eq:Mmdm})  \\         
         $r_1$ & \thead[l]{Distance of an individual MDM in the binary\\ to the COM.} & Above Eq.~\eqref{eq:EOM_Newton}\\
         $\Ueff$ & The effective potential between the MDM binary. & Eq.~\eqref{eq:Ueff}\\
         $r_{1,{\rm max/min}}$ & \thead[l]{The maximum/minimum orbital radii of an\\ individual MDM from the COM. Solved from $E=\Ueff$.} & Below Eq.~\eqref{eq:period_integral}\\
         $\GpoGbarrier$ & \thead[l]{Barrier may exist in $\Ueff$ only when $\GpoG>\GpoGbarrier$.} & Below Eq.~\eqref{eq:dUeff_numerator}\\
         $r_{1,{\rm barrier}}$ & Position of the barrier in $\Ueff$. & Below Eq.~\eqref{eq:dUeff_numerator}\\
         $\Ubarrier$ & Height of the barrier in $\Ueff$. & Below Eq.~\eqref{eq:dUeff_numerator}\\
         $J_{\rm max/min}$ & \thead[l]{The maximal/minimal value of $J$ for $\Ueff$ to possess\\ a barrier when $\GpoG>\GpoGbarrier$.} & Eq.~\eqref{eq:J_max},~\eqref{eq:J_min}\\
         $\Ucross$ & Effective potential when $r_1 = \mmed^{-1}/2$ and $J<J_\text{min}$. & Sec.~\ref{sec:config2} \\         
         $\tau_{\rm GR/mix/\GpoG GR/DF}$ & \thead[l]{The lifetimes of a binary when its initial configuration\\ corresponds to Configuration 1-4, respectively.} & \eqref{eq:tau_GR},~\eqref{eq:tau_mixedrange},~\eqref{eq:tau_betaGR},~\eqref{eq:tau_DF}\\
         $t$ & The cosmic time when the binary merges. & Eq.~\eqref{eq:SGWB} \\
         $\fGW$ & Frequency of the SGWB observed today. & Eq.~(\ref{eq:SGWB}) \\
         $\fGWs$ & \thead[l]{Frequency of the GW emitted at the source.} & Below Eq.~\eqref{eq:dEGWdf_GR}\\
         $f_{\rm trans}$ & \thead[l]{The $\fGWs$ value before the binary evolves to\\ Configuration 3.} & Sec.~\ref{sec:config3}\\
         $f_{\GpoG}$ & \thead[l]{The $\fGWs$ value around which the strength of the DF\\ is activated.} & Sec.~\ref{sec:config3}\\
         $\fmaxs$ & \thead[l]{The maximum GW frequency that can be generated by\\ the binary MDM.} & Sec.~\ref{sec:config4} \\
         $\fdm$ & Fraction of DM energy density made up of MDMs. & Below Eq.~(\ref{eq:rate}) \\
         $\Rdec$ & \thead[l]{Scale factor when two nearest-neighbor MDM objects \\ decouple from the Hubble flow.} & Eq.~(\ref{eq:Rdec}) \\
         $a$ & \thead[l]{Binary semimajor axis (or a similar quantity in the \\ mixed case).} & Eqs.~(\ref{eq:EJtoae}), (\ref{eq:a0_DF}), App.~\ref{sec:params_ae} \\
         $e$ & Binary eccentricity (or a similar quantity in the mixed case). & \thead[c]{Eqs.~(\ref{eq:EJtoae}), (\ref{eq:e0_definition}), App.~\ref{sec:params_ae}} \\
         $\aemit$ & DF mediator emission is expected to start for $a<\aemit$. & Above Eq.~\eqref{eq:dEdf_calc}\\
         $dE_\text{GW}/d\fGWs$ & GW spectrum emitted by an individual binary. & \thead[c]{(\ref{eq:dEGWdf_GR}), (\ref{eq:dEGWdf_betaGR}), (\ref{eq:dEdfGWs_DF}), (\ref{eq:dEdfGWs_trans}), (\ref{eq:dEdfGWs_analytic}) }   \\
         $\Omega_\text{GW}$ & SGWB spectrum. & Eq.~(\ref{eq:SGWB})    
    \end{tabular}

\newpage
\bibliography{DarkBinary}
\bibliographystyle{JHEP}

\end{document}